\documentclass[aps,prb,twocolumn,groupedaddress,showpacs]{revtex4} 

\usepackage{amssymb}
\usepackage{graphicx}
\usepackage{amsmath}
\usepackage{bbm}
\usepackage[colorlinks=true]{hyperref}
\begin{document}

\title{Andreev quantum dot with several conducting channels}

\author{
	I.A.~Sadovskyy$^{1}$,
	G.B.~Lesovik$^{2}$,
	G.~Blatter$^{3}$,
	T.~Jonckheere$^{4}$, and
	T.~Martin$^{4}$
}

\affiliation{
	$^{1}$Rutgers University,
	136 Frelinghuysen Road, Piscataway, New Jersey, 08854, USA
}

\affiliation{
	$^{2}$L.D.~Landau Institute for Theoretical Physics RAS,
	Akademika Semenova av., 1-A, 142432 Chernogolovka, Russia
}

\affiliation{
	$^{3}$Theoretische Physik, Wolfgang-Pauli-Str. 27, ETH-Zurich, CH-8093 
	Z\"urich, Switzerland
}

\affiliation{
	$^{4}$Centre de Physique Th\'eorique, CNRS UMR 7332, 
	Aix-Marseille Universit\'e, Case 907, F-13288 Marseille, France
}

\date{\today}

\begin{abstract} 
We study an Andreev quantum dot, a quantum dot
inserted in a superconducting ring, with several levels or conducting
channels. We analyze the degeneracy of the ground state as a function of the
phase difference and of the gate voltage and find its dependence on the
Coulomb interaction within and between channels. We compute a (noninteger)
charge of the dot region and Josephson current. The charge-to-phase and
current-to-gate voltage sensitivities are studied. We find that, even in the
presence of Coulomb interaction between the channels, the sensitivity
increases with the number of channels, although it does not scale linearly as
in the case with no interactions. The Andreev quantum dot may therefore be
used as a sensitive detector of magnetic flux or as a Josephson transistor.
\end{abstract}


\maketitle

\section{Introduction
\label{sec:introduction}}

In the middle of the last century, Josephson showed that a nondissipative
current can flow between two superconductors separated by an intermediate
region when a phase difference is applied between the two
superconductors.\cite{Josephson:1962,Gennes:1964,Anderson:1963} This
intermediate region can, in practice, be composed of an insulator, a normal
metal, a constriction, etc. Recent developments of nanotechnology allow us to
insert quantum dots, e.g., using carbon nanotubes
\cite{JarilloHerrero:2006,Cleuziou:2006} in this region. These types of
Josephson junctions are called Andreev quantum
dots.\cite{Chtchelkatchev:2003,Sadovskyy:2007} The charge of the Andreev dot
was shown to vary smoothly with the phase difference between the
superconductors\cite{Sadovskyy:2007,Engstrom:2004} in the absence of Coulomb
interaction.

Reference~\onlinecite{Sadovskyy:2007} showed that the charge could be tuned
continuously from~$0$ to~$2e$ depending on the phase difference~$\varphi$ and
gate voltage~$V_{\rm g}$. In Refs.~\onlinecite{Rozhkov:2000} and 
\onlinecite{Sadovskyy:2007b} it was shown that the Coulomb interaction can change 
the property of the ground state, from a usual nondegenerate state with two
quasiparticles to a doubly degenerate state with one quasiparticle. The
present article is a logical continuation of the above-mentioned work and is
devoted to the case of several channels. These channels may appear due to the
presence of two orbits in a single-wall carbon nanotubes and in multiwall
nanotubes\cite{Saito:1998}, or due to an inhomogeneous distribution of
transverse quantized energy levels in metal wires.\cite{Falko:1992} In this
article, we answer the natural questions: How is the charge dependence on the
phase difference affected by the presence of such channels, what is the effect
of Coulomb interaction between channels, and how does this dependence scale
with the number of channels? The calculation is done in the limit of a large
superconducting gap.

\begin{figure}[b]
	\includegraphics[width=5.6cm]{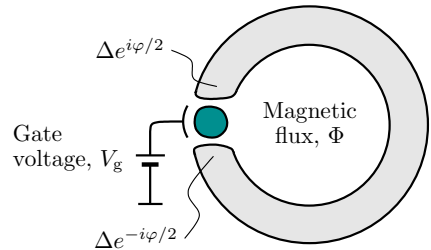}
	\caption{
(Color online) Josephson junction inserted into a superconducting loop which is driven by an external magnetic flux~$\Phi$ and gate voltage~$V_{\rm g}$.
	}
	\label{fig:loop}
\end{figure}

We describe the ground state properties for different setups with different
strengths of the regular Andreev reflection (AR)\cite{Andreev:1964a} and
crossed Andreev reflection
(CAR)\cite{Deutscher:2000,Beckmann:2004,Beckmann:2006,Lesovik:2001}
for different types of Coulomb interaction, such as metal nanowires with
$1$ channel, single-wall carbon nanotubes (SWNT) with $2$ orbital channels,
multiwall carbon nanotubes (MWNT) with $4$ channels, and two separate SWNTs
with $2\times 2$ channels. In this article, we mainly focus on the dependencies
on the tunable parameters gate voltage and superconducting phase difference. We
consider the interplay between phase-sensitive singlet (nondegenerate) and
phase-insensitive doubled (doubly degenerate) states from different channels,
study the current-to-gate voltage and charge-to-phase sensitivities, and discuss
two applications, the Josephson transistor\cite{Kuhn:2001} and a magnetometer based
on an Andreev quantum dot.\cite{Sadovskyy:2007b}

In Sec.~\ref{sec:setup}, we describe a setup based on a quantum dot with
several levels, or alternatively an intermediate region composed of several
single-wall nanotubes/a multiwall nanotube. The model Hamiltonian is
introduced in Sec.~\ref{sec:model}. In Sec.~\ref{sec:scheme}, we represent the
iterative scheme providing the matrix elements of the $N$-channel Hamiltonian.
Using these results, we study the degeneracy and energy of the ground state as
a function of the superconducting phase difference and gate voltage in
Sec.~\ref{sec:degeneracies}. Furthermore, we analyze the magnetic properties of
the dot in Sec.~\ref{sec:magnetic}. The charge on the dot and the Josephson
current are calculated in Sec.~\ref{sec:charge_current}, and their ``sensitivities'' are
discussed in Sec.~\ref{sec:sensitivity}.

\section{Setup
\label{sec:setup}}

As shown in Fig.~\ref{fig:loop}, the setup consists of a structured Josephson
junction in the form of an Andreev quantum dot inserted into a superconducting
loop. We consider a quantum dot with one or several conducting channels and
account for the Coulomb interaction and crossed Andreev reflection. In the
Andreev dot without Coulomb interactions, strong normal reflection 
(electron$\,\to\,$electron, hole$\,\to\,$hole) competes with Andreev reflection
(electron$\,\to\,$hole, hole$\,\to\,$electron), and as a result a strong
dependence of the charge and current on the superconducting phase difference
occurs. Every conducting channel contributes to the
charge\cite{Sadovskyy:2007} and current.\cite{Chtchelkatchev:2000,Kuhn:2001}
Our Andreev dots describe a situation where the resonance width $\Gamma$ is
much smaller than the superconducting gap $\Delta$ ($\Gamma \ll \Delta$) and
the length of the normal part $L$ is much shorter than the superconducting
coherence length $\xi$ ($L \ll \xi$). The superconducting phase difference
$\varphi$ across the junction is controlled by the magnetic flux $\Phi$
passing through the ring: $\varphi = 2\pi \Phi/\Phi_0$, where $\Phi_0$ is the
flux quantum $\Phi_0 = 2|e|/2\pi\hbar$ and $e=-|e|$ is a charge of one
electron.

Realistic experimental setups might be based on a single-wall
nanotube,\cite{JarilloHerrero:2006,Cleuziou:2006} on two or more single-wall
nanotubes placed in parallel, or on a multiwall nanotube. The case with two
parallel single-wall nanotubes is illustrated in 
Fig.~\hyperref[fig:andreev_dot]{\ref{fig:andreev_dot}(b)}:
This case is interesting because the Coulomb interaction between nanotubes can
be smaller than in each nanotube. If such nanotubes are separated by a large
distance, the effects of the tubes will be additive. The second interesting and
experimentally realized case is the multiwall nanotube with approximately
equal interactions in each channel and between them, as depicted in
Fig.~\hyperref[fig:andreev_dot]{\ref{fig:andreev_dot}(c)}.

\section{Model
\label{sec:model}}

\begin{figure}[t]
	\includegraphics[width=0.76\linewidth]{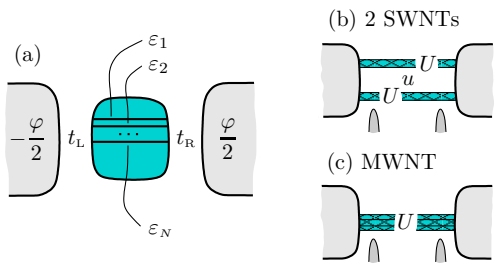}
	\caption{
(Color online) Structure of the Andreev quantum dot with~$N$ channels. (a)~A dot with~$N$
normal levels with corresponding energies $\varepsilon_1, \ldots,
\varepsilon_{\scriptscriptstyle N}$ is inserted between two superconductors
with phase difference~$\varphi$ across the tunnel junctions
$t_{\rm\scriptscriptstyle L}$ and $t_{\rm\scriptscriptstyle R}$. (b)~Quantum
dot system based on two parallel single-wall nanotubes. The barriers
$t_{\rm\scriptscriptstyle L}$ and $t_{\rm\scriptscriptstyle R}$ are formed by
two external gates. The Coulomb interaction $u$ between the channels may be
smaller than the interaction $U$ inside a given channel. (c)~Quantum dot
based on a multiwall nanotube. The Coulomb energy between orbital channels is
approximately equal to the ones inside each channel.
	}
	\label{fig:andreev_dot}
\end{figure}

We consider a SWNT suspended between two superconducting leads,
cf.~Fig.~\hyperref[fig:andreev_dot]{\ref{fig:andreev_dot}(a)}. Additional gates 
placed above the nanotube allow us to define precisely the extent of the 
quantum dot and therefore allow us to modulate its energy levels. An overall 
gate voltage allows us to apply an electric field to the entire structure and 
changes the position of the normal dot levels $\varepsilon_i$ (index 
$i = 1, \ldots, N$ labels the channels). The normal island(s) can be described 
as zero dimensional objects.

The model can be described by a Hamiltonian which includes the dot and its
internal degrees of freedom, the leads, and the tunnel coupling between the
latter two, ${\hat H} = {\hat H}_{\rm\scriptscriptstyle D} + {\hat
H}_{\rm\scriptscriptstyle S} + {\hat H}_{\rm\scriptscriptstyle T}$. The first
part ${\hat H}_{\rm\scriptscriptstyle D}$ describes the quantum dot with $N$
normal levels
\begin{equation}
	{\hat H}_{\rm\scriptscriptstyle D} =
	\sum\limits_{i=1}^N \varepsilon_i {\hat n}_i +
	\sum\limits_{i,j=1}^N U_{ij} {\hat n}_i {\hat n}_j,
	\label{H_dot}
\end{equation}
where the first term represents the interaction with the external gate, ${\hat
n}_i = {\hat n}_{i\uparrow} + {\hat n}_{i\downarrow}$, ${\hat n}_{i\sigma} =
{\hat d}_{i\sigma}^\dag {\hat d}_{i\sigma}^{\phantom\dag}$, and ${\hat
d}_{i\sigma}^\dag$ and ${\hat d}_{i\sigma}^{\phantom\dag}$ are electronic
creation and annihilation operators for the $i^\text{th}$ level in the dot, 
respectively; $\sigma = \uparrow, \downarrow$, and~$\varepsilon_i$ is the 
$i^\text{th}$ energy level with respect to the Fermi level.
The second term describes the Coulomb interaction within the channels
($U_{ii}$) and between different channels ($U_{ij}$, $i\neq j$). The symmetric
matrix $U_{ij}$ is positive definite, and all its elements are
positive.\cite{Landau:1984}
The lead Hamiltonian describes two BCS superconductors [with a lead index
$\ell = \rm L, R$ (left, right)]
\begin{equation}
	{\hat H}_{\rm\scriptscriptstyle S} =
	\sum\limits_{\ell,k} {\hat \Psi}^\dag_{\ell,k}
	(\xi_k{\hat \sigma}_z + \Delta{\hat \sigma}_x)
	{\hat \Psi}^{\phantom\dag}_{\ell,k}, \quad
	{\hat \Psi}_{\ell,k} = \left[\!\! \begin{array}{l}
		\psi^{\phantom\dag}_{\ell,k,\uparrow} \\
		\psi^\dag_{\ell,-k,\downarrow}
	\end{array} \!\!\right] \!,
	\label{H_superconductors}
\end{equation}
with an energy dispersion in the superconducting leads $\xi_k = \hbar^2 k^2/2m -
E_{\rm\scriptscriptstyle F}$ and an absolute value of the gap $\Delta$ in the
bulk of the superconductors. The electron hopping term between dots and leads
is given by
\begin{equation}
	{\hat H}_{\rm\scriptscriptstyle T} =
	\sum\limits_{\ell,k} \big( {\hat \Psi}^\dag_{\ell,k} 
		\mathcal{\hat T}_\ell {\hat d} + {\rm H.c.} \big), \quad
	{\hat d} = \left[\! \begin{array}{l}
		{\hat d}_\uparrow \\
		{\hat d}_\downarrow^\dag
	\end{array} \!\right] \!,
	\label{H_tunneling}
\end{equation}
where $\mathcal{\hat T}_{\rm\scriptscriptstyle L,R} = t_{\rm\scriptscriptstyle
L,R} {\hat \sigma}_z e^{\pm i{\hat \sigma}_z \varphi/4}$ and $t_\ell$ are
tunneling amplitudes between the superconductors and the dot. The superconductor
has a phase $\varphi/2$ on the left and $-\varphi/2$ on the right. The calculation
of observables for a thermal equilibrium system starts from the
evaluation of the partition function $Z = {\rm Tr}\{e^{-\beta {\hat
H}}\}$, where $\beta$ is the inverse temperature.

We work in the limit of a large superconducting gap $|\varepsilon_i|$,
$U_{ij}$, $\Gamma_{ij} \ll \Delta$ (the so-called $\Delta \to \infty$ limit),
where~$\Gamma_{ij}$ are the resonance half-widths originating from tunneling
processes [Eq.~(\ref{H_tunneling})]. In this limit, one can integrate over the lead
degrees of freedom and benefit from the absence of retardation effects due to
the latter. The Hamiltonian $\hat H$ can be rewritten in a simpler form
\begin{align}
	\mathcal{\hat H}_{\scriptscriptstyle N} = &
	\sum\limits_{i=1}^N \varepsilon_i(V_{\rm g}) {\hat n}_i
	+ \sum\limits_{i,j=1}^N U_{ij} {\hat n}_i {\hat n}_j \nonumber \\
	+ & \sum\limits_{i,j=1}^N {\tilde\Gamma}_{ij}(\varphi) 
	[{\hat d}_{i\downarrow} {\hat d}_{j\uparrow} + {\rm H.c.}],
	\label{H_eff}
\end{align}
describing the dot alone with the superconductors defining the boundary
conditions.\cite{Zazunov:2006,Sadovskyy:2010,MartinRodero:2011} We explicitly indicate the
number of channels $N$ in the Hamiltonian $\mathcal{\hat
H}_{\scriptscriptstyle N}$ [Eq.~(\ref{H_eff})] and build an iterative scheme
for its matrix elements $\mathcal{H}_{\scriptscriptstyle N}$ on $N$.
Upon integration, Eqs.~(\ref{H_superconductors}) and~(\ref{H_tunneling})
generate the last term in Eq.~(\ref{H_eff}), which implies that this
Hamiltonian does not conserve the number of electrons. The coefficients
${\tilde\Gamma}_{ij}$ are directly derived from the tunneling amplitudes
$t_{\rm\scriptscriptstyle L,R}$ and describe two-particle tunneling processes
involving both Andreev and normal scattering events at the left and right
boundaries. For symmetric boundaries with equal transparencies
$|t_{\rm\scriptscriptstyle L}|^2 = |t_{\rm\scriptscriptstyle R}|^2 = |t|^2$, we
find $\tilde\Gamma_{ij} = \Gamma_{ij} \cos(\varphi/2)$. The symmetric matrix
$\Gamma_{ij}$ describes Andreev reflection, the annihilation of two
quasiparticles from the $i^\text{th}$ and $j^\text{th}$ channel ${\hat
d}_{i\downarrow} {\hat d}_{j\uparrow}$ (with simultaneous creation of a Cooper
pair in the superconductor) and the opposite process. The diagonal terms
${\tilde\Gamma}_{ii}$ correspond to the regular Andreev reflection inside
the $i^\text{th}$ channel, the off-diagonal terms ${\tilde\Gamma}_{ij}$, ($i\neq
j$)~ describes the crossed Andreev reflection (CAR) between the $i^\text{th}$
and $j^\text{th}$ channel. The element $\Gamma_{ij}$ can be treated as the
resonance half-width of the normal tunneling process between the $i^\text{th}$
and $j^\text{th}$ channel. Note that for a diagonal matrix $\Gamma_{ij}$ and
$U_{ij}$ (with $\Gamma_{ij}=0$, $U_{ij}=0$, $i\neq j$) the channels are
additive and the answers for all physical quantities can be obtained by
summation over all channels.

\section{Iterative scheme
\label{sec:scheme}}

The Hamiltonian Eq.~(\ref{H_eff}) for a single channel (the first channel in
the iterative scheme presented below) has the following form
\begin{equation}
	{\hat H}_1 = 
	\varepsilon_1 {\hat n}_1
	+ U_1 {\hat n}_1^2
	+ \tilde\Gamma_1 [{\hat d}_{1\downarrow} {\hat d}_{1\uparrow} + {\rm H.c.}],
	\label{H_eff_1}
\end{equation}
where the $\hat{n}_{1\sigma}^2$ contributions from the sum ${\hat n}_1^2 =
(\hat{n}_{1\uparrow} + \hat{n}_{1\downarrow})^2$ has been absorbed in a shift
of~$\varepsilon_1$. We denote the Hamiltonian of the $i^\text{th}$ channel as
${\hat H}_i$ and the total Hamiltonian of $N$ channels as $\mathcal{\hat
H}_{\scriptscriptstyle N}$. The corresponding matrix of the $H_1$ Hamiltonian
has dimensions $4\times 4$ and can be calculated in a basis of four states
$|\nu\rangle_1 = \{|0\rangle_1$, $|\!\!\uparrow\rangle_1$,
$|\!\!\downarrow\rangle_1$, $|2\rangle_1\}$, the state with no electrons
$|0\rangle$, with one electron with spin up $|\!\!\uparrow\rangle = {\hat
d}_\uparrow^\dag |0\rangle$ or spin down $|\!\!\downarrow\rangle = {\hat
d}_\downarrow^\dag |0\rangle$, and the two electron state $|2\rangle = {\hat
d}_\uparrow^\dag {\hat d}_\downarrow^\dag |0\rangle$
\begin{align}
	H_1 & =
	\varepsilon_1 n_1 + U_1 n_1^2 + 
	{\rm adiag}_1 \{ {\tilde\Gamma}_1, 0, 0, {\tilde\Gamma}_1 \}
	\nonumber \\ 
	& = \left[ \!
	\begin{array}{cccc}
		0 & 0 & 0 & {\tilde\Gamma}_1 \\
		0 & \varepsilon_1 + U_1 & 0 & 0 \\
		0 & 0 & \varepsilon_1 + U_1 & 0 \\
		{\tilde\Gamma}_1 & 0 & 0 & 2\varepsilon_1 + 4U_1
	\end{array}
	\! \right]_1 \!,
	\label{H_eff_1_me}
\end{align}
where 
\begin{equation}
	n_1 = 
	n_{1\uparrow} + n_{1\downarrow} =
	Q_1/e = 
	{\rm diag}_1 \{0, 1, 1, 2\}
	\label{Q_op}
\end{equation}
are matrix elements of the dimensionless charge operator and ${\hat Q}_1 = e {\hat n}_1$.
In what follows, the expression ${\rm adiag}\{x_1, x_2, \ldots,
x_{\scriptscriptstyle N}\}$ stands for the $N\times N$ matrix $A_{ij}$, where
$A_{i,N-i+1}=x_i$ and $A_{i,j}=0$ for $j\neq N-i+1$. The eigenvalues of
Eq.~(\ref{H_eff_1_me}) can be easily calculated. They consist of two
nondegenerate levels with energies
\begin{equation}
	E_{0/2} = \varepsilon_1 + 2U_1 \mp 
	\sqrt{(\varepsilon_1 + 2U_1)^2 + {\tilde\Gamma}_1^2}
	\label{S_en}
\end{equation}
and a doubly degenerate level with energy
\begin{equation}
	E_1 = \varepsilon_1 + U_1
	\label{D_en}
\end{equation}
which can be split into two separate levels by an external magnetic field
through the Zeeman effect. Usually, the ground state of the Hamiltonian
Eq.~(\ref{H_eff_1_me}) is given by the singlet $E_0$. But in the case of
nonzero Coulomb interaction $U_1 > 0$, there exists a range of parameters
$\varphi$ and $\varepsilon_i$ for which the ground state is doubly degenerate
with energy $E_1$. The doublet region is defined by the inequality
\begin{equation}
	(\varepsilon_1 + 2U_1)^2 + {\tilde\Gamma}_1^2 < U_1^2,
	\label{D_reg}
\end{equation}
where ${\tilde\varepsilon} = \varepsilon_1 + 2U_1$. For an asymmetric dot
($|t_{\rm\scriptscriptstyle L}| \neq |t_{\rm\scriptscriptstyle R}|$), in the
presence of electron-phonon interaction or other perturbations, the ``critical''
$U_1$ is different from zero, $U_{1,\rm\scriptscriptstyle C} >
0$.\cite{Sadovskyy:2007b} In this article, we ignore such nonidealities and
concentrate on the simplest case.
The same results are valid for any other $i>1$ channel; they can be obtained
by the substitution $\varepsilon_1 \to \varepsilon_i$, $\Gamma_1 \to
\Gamma_i$, and $U_1 \to U_i$.

\begin{figure}[tb]
	\includegraphics[width=0.99\linewidth]{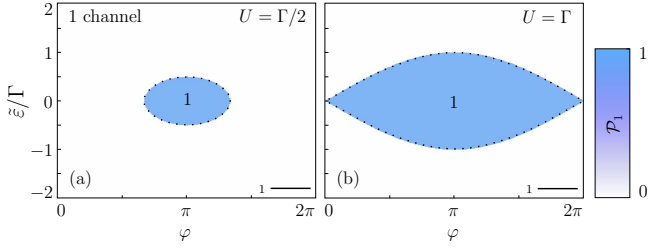}
	\caption{
(Color online) System degeneracy $\mathcal{P}_1$ in the ground state in 
$(\varphi, {\tilde\varepsilon})$ space for the channel $N=1$.
	}
	\label{fig:degeneracies_1ch}
\end{figure}

\begin{figure*}[tb]
	\includegraphics[width=13.3cm]{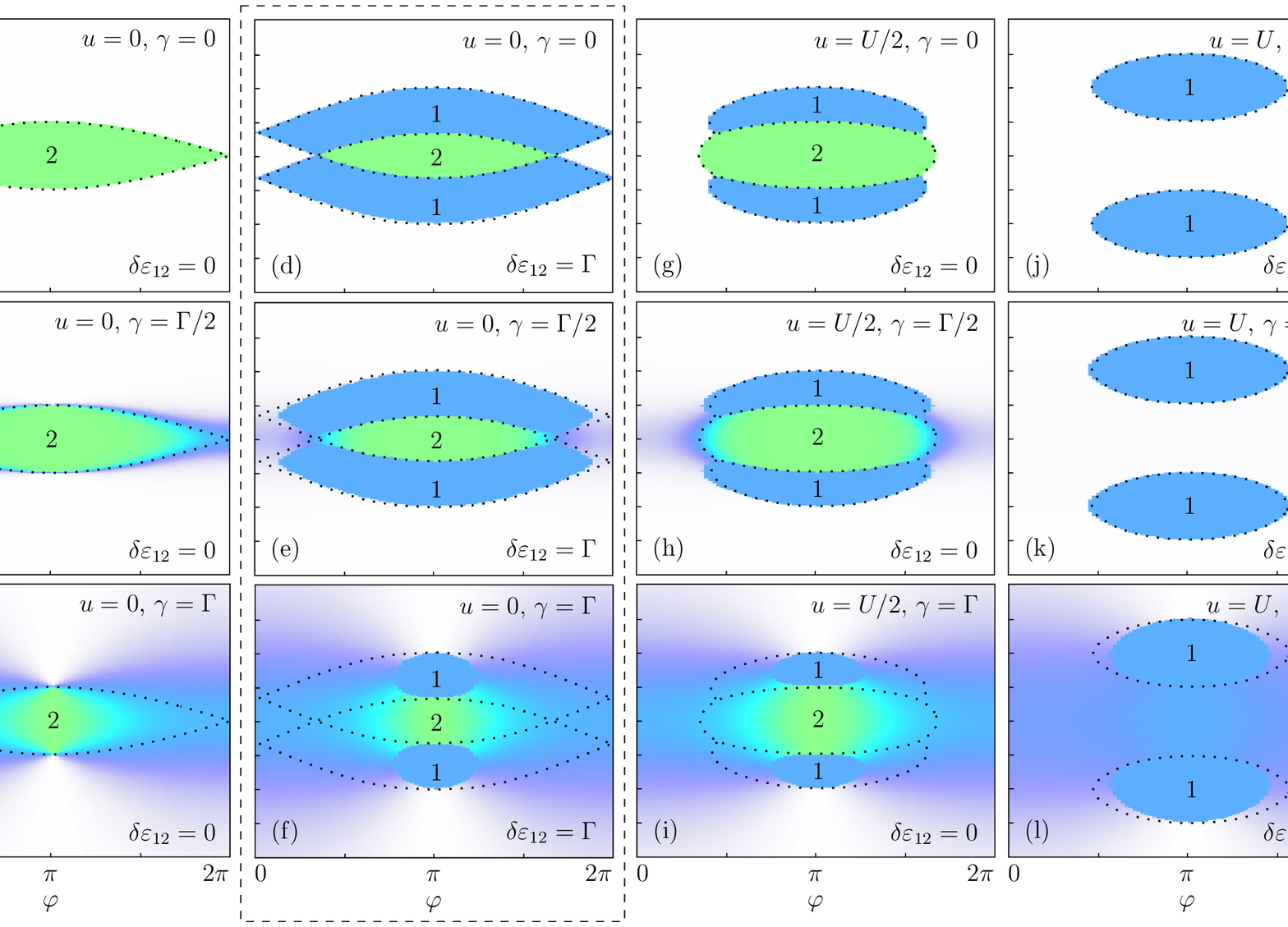}
	\caption{
(Color online) System degeneracy $\mathcal{P}_2$ in the ground state in the $(\varphi,
{\tilde\varepsilon})$ space for $N=2$ channels. The Coulomb matrix $U_{ij}$ is
defined by Eq.~(\ref{U_ij_mod}), CAR is defined by Eq.~(\ref{G_ij_mod}). The
Coulomb interaction inside each channel is equal and equal to the resonance
width, $U = \Gamma$. The Coulomb interaction between channels varies from left
[$u=0$ in (a)-(c)] to right [$u=U$ in (j)-(l)]; The CAR parameter varies
from top [$\gamma=0$ in (a), (d), (g), and (j)] to bottom [$\gamma=\Gamma$ in
(c), (f), (i), and (l)]. The difference in dot levels $\delta\varepsilon_{12}
= \varepsilon_2-\varepsilon_1 = \Gamma$ in (d)-(f) and
$\delta\varepsilon_{12} = 0$ in all other plots. (a)~In the noninteracting
case the degenerate region with $\mathcal{P}_2=2$ (green) inside the
nondegenerate region $\mathcal{P}_2=0$ (white) is defined by the inequality~(\ref{D_reg})
and coincides with the same region of each channel. (b) and (c)~CAR smears
the border and $\mathcal{P}_2$ goes from 0 to 2 continuously. (c)~Two
doubly-degenerate regions (blue) shifted by $\delta\varepsilon_{12}$; the
intersection represents a fourth-fold degenerate region (green). (e) and
(f)~The border is (nonuniformly) smeared due to CAR, starting from the point
where three regions with $\mathcal{P}_2=0$, $1$, and $2$ join together and
ending with a smeared four-fold region. (g)-(i)~The average Coulomb
interaction between the channels $u=U/2$ separates apart the region with
$\mathcal{P}_2=1$ in (a) by the distance $4u$, but in a different manner from
(b); CAR smears the borders of the $\mathcal{P}_2=2$ region and diminishes the
$\mathcal{P}_2=1$ regions. (j)-(l)~The maximal Coulomb interaction between
channels $u=U$ totally splits the $\mathcal{P}_2=2$ region into two regions
with $\mathcal{P}_2=1$ (the distance is still $4u$); CAR slowly increases
$\mathcal{P}_2$ from 0 to 1 and slightly reduces the $\mathcal{P}_2=1$
regions.
	}
	\label{fig:degeneracies_2ch}
\end{figure*}

The structure of the ground state in the one-channel case is presented in
Fig.~\ref{fig:degeneracies_1ch} as a function of phase $\varphi$ and the
``central'' dot-level ${\tilde\varepsilon} = \varepsilon_1 + 2U_1$. The
blue figures in the center represent the doublet state~\ref{D_en}, their
shapes are defined by the inequality~(\ref{D_reg}), the white space around
corresponds to a usual singlet state with energy~(\ref{S_en}). For two
completely separated single-channel dots with the same levels $\varepsilon_1 =
\varepsilon_2$ connected in parallel, the dependence is the same, however, now
the central figure corresponds to a four-fold degenerate situation. 

We now go back to the $N$-channel case. We calculate the matrix elements of
the Hamiltonian Eq.~(\ref{H_eff}) in the basis $|\nu\rangle_1 \otimes
|\nu\rangle_2 \otimes \ldots \otimes |\nu\rangle_{\scriptscriptstyle N} = \{
{|0,\ldots,0\rangle},$ ${|0,\ldots,\uparrow\rangle},$
${|0,\ldots,\downarrow\rangle}, \ldots,$ ${|2,\ldots,\downarrow\rangle},$
${|2,\ldots,2\rangle} \}$, where $|\nu\rangle_i = \{|0\rangle_i$,
$|\!\!\uparrow\rangle_i$, $|\!\!\downarrow\rangle_i$, $|2\rangle_i\}$ are
states of the Hamiltonian for the single channel case [Eq.~(\ref{H_eff_1_me})]
\begin{align} \label{H_eff_mult}
	\!\! \mathcal{\hat H}_{\scriptscriptstyle N} 
	= \sum\limits_{i=1}^N & \mathbbm{1}_1 \otimes \ldots \otimes {\hat H}_i \otimes \ldots \otimes \mathbbm{1}_{\scriptscriptstyle N} \\ 
	+ \sum\limits_{\substack{i,j = 1 \\ i\neq j}}^N & U_{ij} \, \mathbbm{1}_1 \otimes \ldots \otimes {\hat n}_i \otimes \ldots \otimes {\hat n}_j \otimes \ldots \otimes\mathbbm{1}_{\scriptscriptstyle N} \nonumber \\ 
	+ \sum\limits_{\substack{i,j = 1 \\ i\neq j}}^N & {\tilde\Gamma}_{ij} \, \bigl[
		\mathbbm{1}_1 \otimes \ldots \otimes {\hat d}_{i\downarrow} \otimes \ldots \otimes {\hat d}_{j\uparrow} \otimes \ldots \otimes\mathbbm{1}_{\scriptscriptstyle N} \nonumber \\
		& \hspace{2.2mm} + \mathbbm{1}_1 \otimes \ldots \otimes {\hat d}_{j\uparrow}^\dag \otimes \ldots \otimes {\hat d}_{i\downarrow}^\dag \otimes \ldots \otimes\mathbbm{1}_{\scriptscriptstyle N}
	\bigr].
	\nonumber
\end{align}
Here, the first sum takes into account the interactions inside each channel
(given by the direct tensor product of $H_i$'s and unity operators
$\mathbbm{1}_j$ in the other subspaces). The second sum is the interaction between
channels; each term represents the tensor product of the charge operator in the
$i^\text{th}$ channel and in the $j^\text{th}$ channel [as given by
Eq.~(\ref{Q_op})]. Here, $\mathbbm{1}_i$ is the unit matrix in the Hilbert
subspace of the $i^\text{th}$ channel, and $Q_i$ is the matrix characterizing
the $i^\text{th}$ channel charge. The third sum describes CAR; each term
corresponds to the CAR in the $i^\text{th}$ and $j^\text{th}$ channels. The
Coulomb interaction inside each channel and the regular Andreev reflection are
accounted for in the first term.

Let us construct a recursive procedure in $N$ for the matrix elements
corresponding to the $N$-channel Hamiltonian. As the initial step of the
recursion $N=1$ we take the matrix~(\ref{H_eff_1_me}). At the second step
$N=2$ we build a matrix in $|\nu\rangle_1 \otimes |\nu\rangle_2$ in the
following way:
\begin{align}
   \mathcal{H}_2 & =
   {\rm diag}_2 \big\{ 
     H_1,
     H_1 + \varepsilon_2 + U_2 + \mathbb{U}_{21},
     H_1 + \varepsilon_2 + U_2 + \mathbb{U}_{21}, \nonumber \\
     & \phantom{\; = {\rm diag}_2 \big\{} H_1 + 2\varepsilon_2 + 4U_2 + \mathbb{U}_{22}
   \big\} \label{H_eff_2_me} \\
   & + {\tilde\Gamma}_2 \left[ \! \begin{array}{cccc}
     0 & 0 & 0 & \mathbbm{1}_1 \\
     0 & 0 & 0 & 0 \\
     0 & 0 & 0 & 0 \\
     \mathbbm{1}_1 & 0 & 0 & 0
   \end{array} \! \right]_2 
   + {\tilde\Gamma}_{12} \left[ \!\! \begin{array}{cccc}
     0 & \! {\phantom -} d_{1\downarrow}^\dag  & d_{1\uparrow}^\dag & 0 \\
     d_{1\downarrow} &  0  & 0 & \! -d_{1\uparrow}^\dag \\
     d_{1\uparrow} &  0  & 0 & \! {\phantom -} d_{1\downarrow}^\dag \\
     0 & \! -d_{1\uparrow}  & d_{1\downarrow} & 0
   \end{array} \!\!\right]_2 \!,
\nonumber
\end{align}
where we keep in mind that ${\tilde\Gamma}_{12} = {\tilde\Gamma}_{21}$. The matrices
\begin{equation}
   d_{i\uparrow} = 
   \left[ \begin{array}{cccc}
     0 & 0 & 0 & 0 \\ 
     1 & 0 & 0 & 0 \\ 
     0 & 0 & 0 & 0 \\ 
     0 & 0 & 1 & 0    
   \end{array} \right]_i\!, \quad
   d_{i\downarrow} = 
   \left[ \begin{array}{crcc}
     0 &  0  & 0 & 0 \\ 
     0 &  0  & 0 & 0 \\ 
     1 &  0  & 0 & 0 \\ 
     0 & -1  & 0 & 0    
   \end{array} \right]_i
   \nonumber
\end{equation}
and the corresponding Hermitian conjugates $d_{i\uparrow}^\dag$,
$d_{i\downarrow}^\dag$ are the matrix elements of the operators ${\hat
d}_{i\uparrow}^{\phantom\dag}$, ${\hat d}_{i\downarrow}^{\phantom\dag}$ and
${\hat d}_{i\uparrow}^\dag$, ${\hat d}_{i\downarrow}^\dag$ in the
$i^\text{th}$ channel basis. This is a $16\times 16$ matrix (the Hilbert space
of the $1^\text{st}$ channel with dimension~4 should be multiplied by the
Hilbert space of the $2^\text{nd}$ one with the same dimension). Each
``element'' is a $4\times 4$ block, ``diag'' and ``adiag'' to be in the
$2^\text{nd}$ channel subspace. $H_1$ originates from Eq.~(\ref{H_eff_1_me}),
with scalar elements to be multiplied by $\mathbbm{1}_1$. The matrices
$\mathbb{U}_{21} = {\rm diag}_1\{0, 0, 0, 8U_{12}\}$ and~$\mathbb{U}_{22} =
{\rm diag}_1\{0, 8U_{12}, 8U_{12}, 0\}$ are based on the term $U_{12}$ which
is responsible for the interaction between channels. For the noninteracting
case $U_{12}=0$, the structure of the ground state is presented in
Fig.~\hyperref[fig:degeneracies_2ch]{\ref{fig:degeneracies_2ch}(a)} [$\varepsilon_2 - \varepsilon_1 = 0$] and
\hyperref[fig:degeneracies_2ch]{\ref{fig:degeneracies_2ch}(d)} [$\varepsilon_2 - \varepsilon_1 = \Gamma$] in
coordinates $(\varphi, \tilde\varepsilon)$, where ${\tilde\varepsilon} =
(\varepsilon_1 + \varepsilon_2)/2 + U_1 + U_2$ is the average and renormalized
normal level in the dot. The interacting case is shown in
Figs.~\hyperref[fig:degeneracies_2ch]{\ref{fig:degeneracies_2ch}(g)} and \hyperref[fig:degeneracies_2ch]{\ref{fig:degeneracies_2ch}(j)}, where
${\tilde\varepsilon} = (\varepsilon_1 + \varepsilon_2)/2 + U_1 + U_2 +
2U_{12}$; see Sec.~\ref{sec:degeneracies} for more details.

In the same way we can write a recursive formula
for~$\mathcal{H}_{\scriptscriptstyle N}$. The easiest way to write it down is
\begin{align}
	\mathcal{H}_{\scriptscriptstyle N} & =
	\sum\limits_{i=1}^N \varepsilon_i n_i + 
	\sum\limits_{i=1}^N U_i n_i^2 + 
	\sum\limits_{i=1}^N {\tilde\Gamma}_i \, {\rm adiag}_i \{1, 0, 0, 1\} \nonumber \\
	& + \sum\limits_{\substack{i,j = 1 \\ i\neq j}}^N U_{ij} n_i n_j 
	+ \sum\limits_{\substack{i,j = 1 \\ i\neq j}}^N {\tilde\Gamma}_{ij}
	\bigl[ d_{i\downarrow} d_{j\uparrow} + d_{j\uparrow}^\dag d_{i\downarrow}^\dag \bigr].
	\label{H_eff_N_me}
\end{align}
Here each matrix with index $i$ should be written as a matrix of size
$4^i\times 4^i$ (i.e., in the subspace of the $i^\text{th}$ channel). Herein
each element of the matrix associated with the $i^\text{th}$ channel is
multiplied by a unitary matrix of size $4^{i-1}\times 4^{i-1}$ and starts
on column $(c-1)\,2^{i-1}+1$ and on row $(r-1)\,2^{i-1}+1$ of the
total matrix, where $c =1,\ldots,4$ is the column and $r =1,\ldots,4$ is the
row in the original $4\times 4$ matrix of the $i^\text{th}$ channel.
Schematically the recursive procedure for building the $N$-channel matrix can
be represented as (for $N=3$)
\begin{equation}
{\small 
\left[\begin{array}{ccc}
  \left[\begin{array}{ccc}
    \left[\begin{array}{c} \cdot \end{array}\right]_1 & \cdots & \left[\begin{array}{c} \cdot \end{array}\right]_1 \\
    \vdots_{\phantom 1} & & \vdots_{\phantom 1} \\
    \left[\begin{array}{c} \cdot \end{array}\right]_1 & \cdots & \left[\begin{array}{c} \cdot \end{array}\right]_1
  \end{array}\right]_2
  & \cdots &
  \left[\begin{array}{ccc}
    \left[\begin{array}{c} \cdot \end{array}\right]_1 & \cdots & \left[\begin{array}{c} \cdot \end{array}\right]_1 \\
    \vdots_{\phantom 1} & & \vdots_{\phantom 1} \\
    \left[\begin{array}{c} \cdot \end{array}\right]_1 & \cdots & \left[\begin{array}{c} \cdot \end{array}\right]_1
  \end{array}\right]_2
  \\
  \vdots_{\phantom 2} & & \vdots_{\phantom 2} \\
  \left[\begin{array}{ccc}
    \left[\begin{array}{c} \cdot \end{array}\right]_1 & \cdots & \left[\begin{array}{c} \cdot \end{array}\right]_1 \\
    \vdots_{\phantom 1} & & \vdots_{\phantom 1} \\
    \left[\begin{array}{c} \cdot \end{array}\right]_1 & \cdots & \left[\begin{array}{c} \cdot \end{array}\right]_1
  \end{array}\right]_2
  & \cdots &
  \left[\begin{array}{ccc}
    \left[\begin{array}{c} \cdot \end{array}\right]_1 & \cdots & \left[\begin{array}{c} \cdot \end{array}\right]_1 \\
    \vdots_{\phantom 1} & & \vdots_{\phantom 1} \\
    \left[\begin{array}{c} \cdot \end{array}\right]_1 & \cdots & \left[\begin{array}{c} \cdot \end{array}\right]_1
  \end{array}\right]_2
\end{array}\right]_3
}
\nonumber
\end{equation}
The smallest ``rectangle'' corresponds to the Hamiltonian $H_1$. In a second
step, we insert such blocks in the Hamiltonian $H_2$ and continue the procedure
till $N$. At each $i^\text{th}$ step, we multiply the current space by the
Hilbert subspace of the $i^\text{th}$ Andreev level.

Note that the same procedure can be realized by using the Bogoliubov-de Gennes
equations and the scattering matrix approach in the tunneling regime (with a
transparency of each junction much less than unity) as was done in
Ref.~\onlinecite{Sadovskyy:2007b} for a single channel. However, this approach
is poorly scalable with $N$ and the calculation is much more cumbersome. See
also Ref.~\onlinecite{Kubala:2003} for the correspondence between the
tunneling Hamiltonian method and the scattering matrix approach.

In the present notation, the first, the second, and the fourth sum in
Eq.~(\ref{H_eff_N_me}) are diagonal. In the absence of superconductivity they
define energy levels of the system, and the Coulomb interaction leads to level
``repulsion.'' The third term of Eq.~(\ref{H_eff_N_me}) includes the
dependence on the superconducting phase difference and mixes the states due to
the presence of Cooper pairs in the superconductors. At $\varphi=\pi$ the
Hamiltonian Eq.~(\ref{H_eff_N_me}) is diagonal and its eigenvalues can be
found analytically.

\section{Energy levels and degeneracies
\label{sec:degeneracies}}

\begin{figure}[b]
	\includegraphics[height=1.9cm]{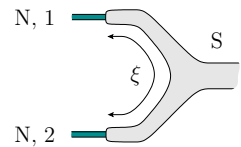}
	\caption{
(Color online) The Crossed Andreev reflection between channels 1 and 2 in the fork (or Y)
geometry is suppressed if the length between the NS interfaces is much larger then
the coherence length $\xi$.
	}
	\label{fig:fork}
\end{figure}

\begin{figure*}[tb]
	\includegraphics[width=13.3cm]{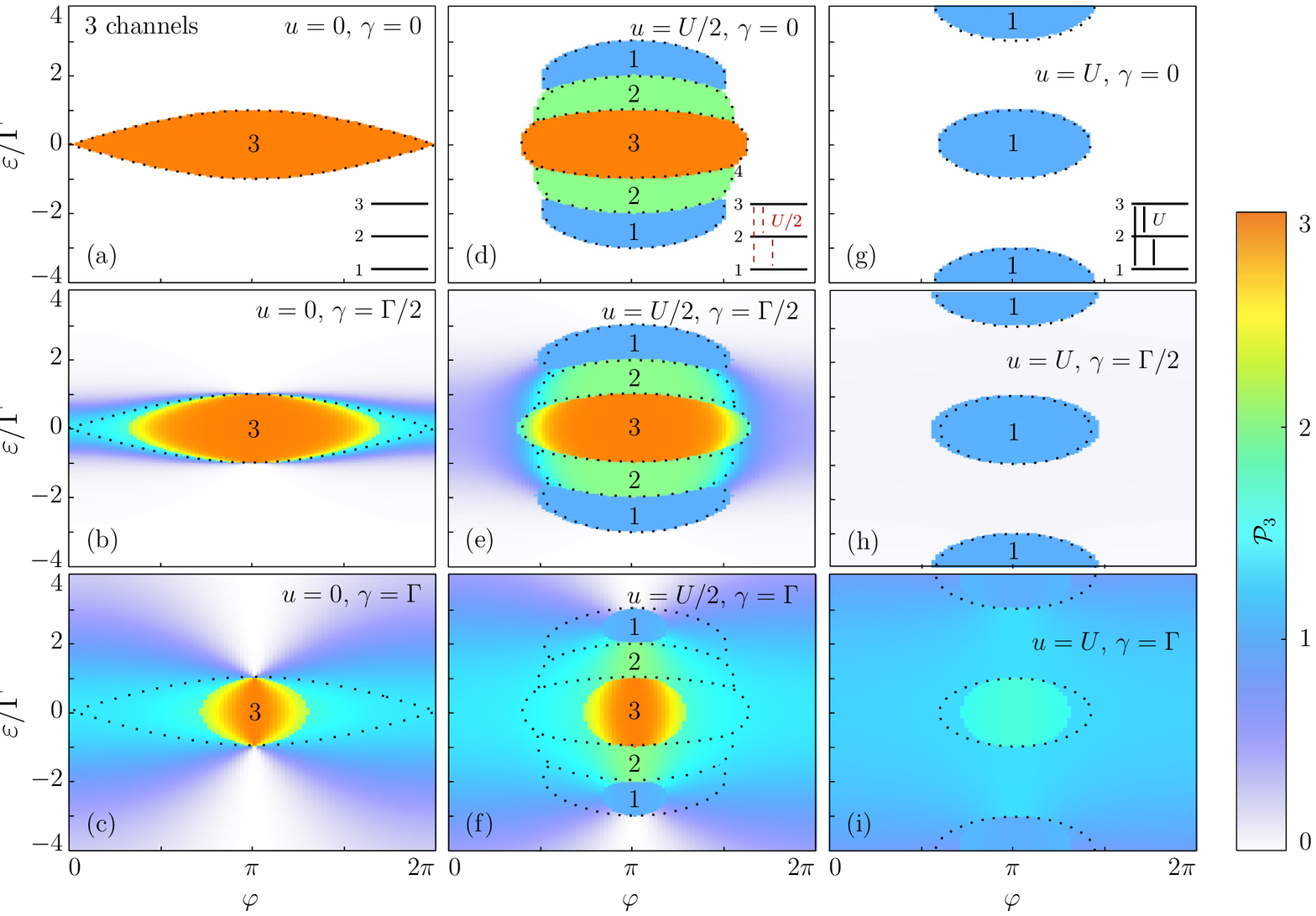}
	\caption{
(Color online) System degeneracy $\mathcal{P}_3$ of the ground state for $N=3$ channels; the
same as in Figs.~\hyperref[fig:degeneracies_2ch]{\ref{fig:degeneracies_2ch}(a)}-\hyperref[fig:degeneracies_2ch]{\ref{fig:degeneracies_2ch}(c)},
\hyperref[fig:degeneracies_2ch]{\ref{fig:degeneracies_2ch}(g)}-\hyperref[fig:degeneracies_2ch]{\ref{fig:degeneracies_2ch}(i)}, and
\hyperref[fig:degeneracies_2ch]{\ref{fig:degeneracies_2ch}(j)}-\hyperref[fig:degeneracies_2ch]{\ref{fig:degeneracies_2ch}(l)}.
	}
	\label{fig:degeneracies_3ch}
\end{figure*}

\begin{figure*}[tb]
	\includegraphics[width=13.3cm]{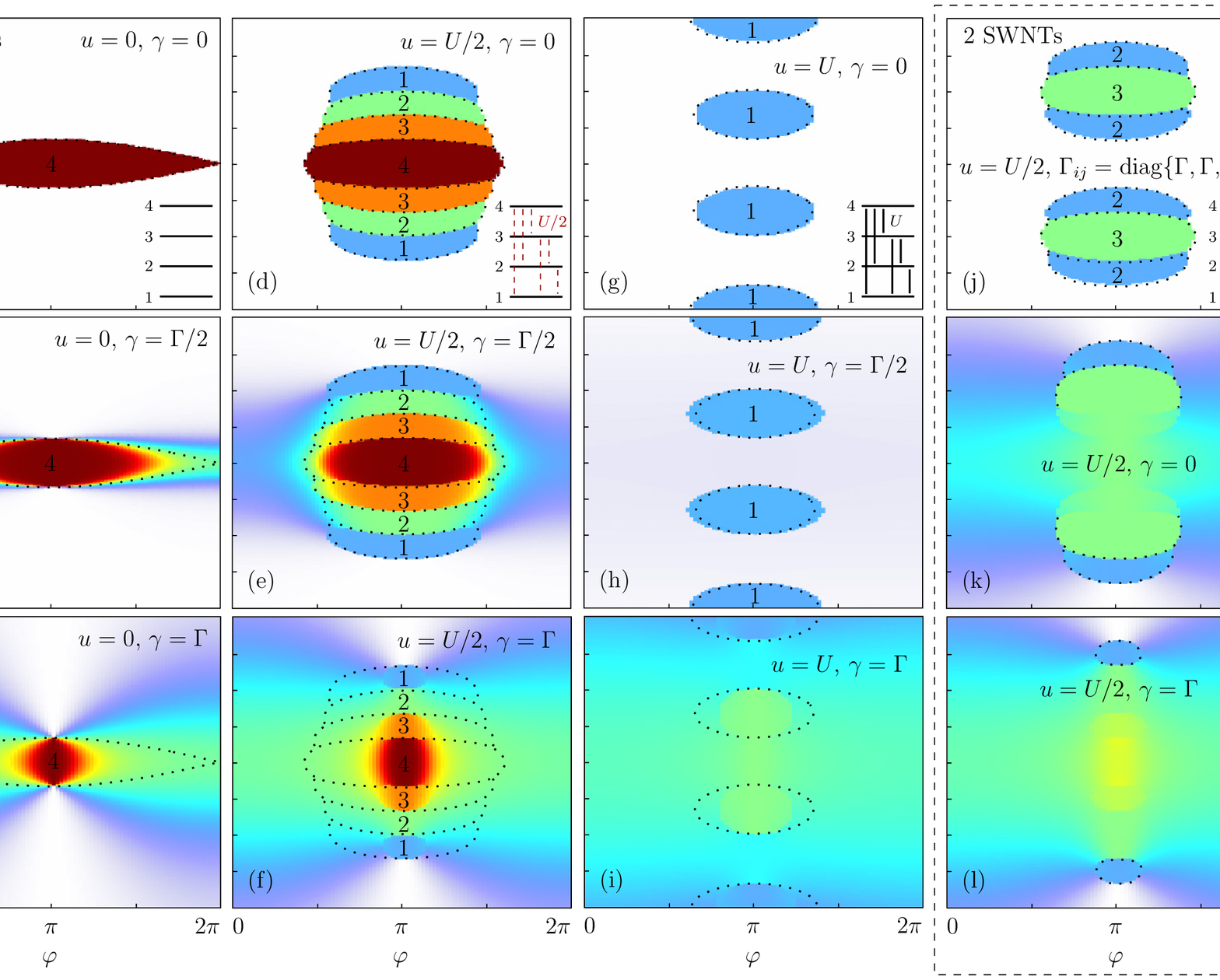}
	\caption{
(Color online) System degeneracy $\mathcal{P}_4$ of the ground state for $N=4$ channels.
(a)-(c)~$\mathcal{P}_4$ goes from 0 to $4$ in a step-like manner for
$\gamma=0$ and smoothly for $\gamma>0$ [compare with
Figs.~\hyperref[fig:degeneracies_2ch]{\ref{fig:degeneracies_2ch}(a)}-\hyperref[fig:degeneracies_2ch]{\ref{fig:degeneracies_2ch}(c)}
and~\hyperref[fig:degeneracies_3ch]{\ref{fig:degeneracies_3ch}(a)}-\hyperref[fig:degeneracies_3ch]{\ref{fig:degeneracies_3ch}(c)}]; the Coulomb matrix $U_{ij}$ is
defined by Eq.~(\ref{U_ij_mod}) and the AR $\Gamma_{ij}$ is defined by
Eq.~(\ref{G_ij_mod}). (d)-(f)~ The average Coulomb interaction between
channels $u=U/2$ transforms the $\mathcal{P}_4=4$ region to $\mathcal{P}_4=1$,
$2$, $3$, and $4$ regions [compare with
Figs.~\hyperref[fig:degeneracies_2ch]{\ref{fig:degeneracies_2ch}(g)}-\hyperref[fig:degeneracies_2ch]{\ref{fig:degeneracies_2ch}(i)}
and~\hyperref[fig:degeneracies_3ch]{\ref{fig:degeneracies_3ch}(d)}-\hyperref[fig:degeneracies_3ch]{\ref{fig:degeneracies_3ch}(f)}]. (g)-(i)~The maximal Coulomb
interaction between channels $u=U$ splits the $\mathcal{P}_4=4$ region to $4$
regions with $\mathcal{P}_4=1$ at a distance $4u$ [compare with
Figs.~\hyperref[fig:degeneracies_2ch]{\ref{fig:degeneracies_2ch}(j)}-\hyperref[fig:degeneracies_2ch]{\ref{fig:degeneracies_2ch}(l)}
and~\hyperref[fig:degeneracies_3ch]{\ref{fig:degeneracies_3ch}(g)}-\hyperref[fig:degeneracies_3ch]{\ref{fig:degeneracies_3ch}(i)}]. (j)~Two parallel SWNT: $U_{ij}$ is
defined by Eq.~(\ref{U_ij_2SWNT}), $\Gamma_{ij}$ by Eq.~(\ref{G_ij_mod}) with
$\gamma=0$. The structure is similar to the double structure in
Fig.~\hyperref[fig:degeneracies_2ch]{\ref{fig:degeneracies_2ch}(g)}. (k) and (l)~Two SWNT: $U_{ij}$ defined by
Eq.~(\ref{U_ij_2SWNT}), $\Gamma_{ij}$ by Eq.~(\ref{G_ij_2SWNT}) for $\gamma=0$
and $\gamma=\Gamma$.
	}
	\label{fig:degeneracies_4ch}
\end{figure*}

In the single channel case, the ground state is formed by nondegenerate
[Eq.~(\ref{S_en})] or doubly degenerate [Eq.~(\ref{D_en})] levels. However,
for a large number of channels one can expect to observe
$2^{\mathcal{P}_{\scriptscriptstyle N}}$ degeneracy, where
$\mathcal{P}_{\scriptscriptstyle N}$ is the number of Andreev levels in the
doublet state.

To characterize the state of each channel let us first determine the operator
\begin{equation}
	{\hat p}_i = {\rm diag}_i \{0, 1, 1, 0\}
	\label{deg_ch}
\end{equation}
which is defined in the Hilbert space of the $i^\text{th}$ channel. For the
case with no interaction, this operator defines the $p_i$ value equal to
$0$ if the $i^\text{th}$ channel is in the singlet state and $1$ if the
$i^\text{th}$ channel is in the doublet state. In the case including 
interaction, $p_i$ may vary from $0$ to $1$. The
$\mathcal{P}_{\scriptscriptstyle N}$ value of the operator
\begin{equation}
	\mathcal{\hat P}_{\scriptscriptstyle N} = 
	\sum\limits_{i=1}^N
		\mathbbm{1}_1 \otimes \ldots \otimes {\hat p}_i \otimes \ldots
		\otimes \mathbbm{1}_{\scriptscriptstyle N}
	\label{deg_tot}
\end{equation}
characterizes the state of the entire system, $0 \leqslant
\mathcal{P}_{\scriptscriptstyle N} \leqslant N$. In the noninteracting case
$\mathcal{P}_{\scriptscriptstyle N}$ gives the number of channels in 
doublet states and can be considered as the degeneracy of the ground state.

In addition, the mechanism for intrastate transitions with different
$\mathcal{P}_{\scriptscriptstyle N}$ should be discussed. As mentioned above
these transitions qualitatively change the properties of the system, the
charging effects, the transport properties, and the magnetic response to the
external field. Such transitions occur, for instance, due to electron-phonon
interactions involving a continuous spectrum.\cite{Sadovskyy:2007} Their rate
is suppressed by the exponential factor $e^{-\Delta/k_{\rm\scriptscriptstyle
B}\Theta}$, where $k_{\rm\scriptscriptstyle B}$ and $\Theta$ are the Boltzmann
constant and temperature, respectively.\cite{Chtchelkatchev:2003,Sadovskyy:2007}
Thus, transitions where the system's spin changes by $1/2$ are more rare than 
transitions where the spin remains unchanged.
Therefore, the above picture for conversion between singlet and
doublet in one channel is valid for an adiabatically slow change of the
parameters $\varphi$ and ${\tilde\varepsilon}$.

Let us consider cases defined by different matrices $U_{ij}$ and
$\Gamma_{ij}$. The diagonal elements of the Coulomb matrix $U_{ij}$ define the
interaction in each channel. The off-diagonal elements describing the
interactions between channels can be of the order of the diagonal elements for
the multichannel wire and about zero for a few separated wires. We can vary
the ``ratio'' between off-diagonal to diagonal elements from 0 to 1. In the
same way, we can vary the off-diagonal elements of $\Gamma_{ij}$, which are of
the order of the diagonal elements and about zero for separated channels,
e.g., see the superconductor in fork geometry in Fig.~\ref{fig:fork}.

\subsection{Toy model}

We parametrize the Coulomb interaction matrix with two parameters
\begin{equation}
	U_{ij} =
	\left\{ \! \begin{array}{ll}
		U, & i=j, \\
		u, & i\neq j. \\
	\end{array} \right.
	\label{U_ij_mod}
\end{equation}
Here $U$ is the interaction within each channel and $u$ is the interaction
between any pair of channels, with $u \leqslant U$. The Coulomb interaction
Eq.~(\ref{U_ij_mod}) describes $N$ parallel wires, each with one channel (no
orbital degeneracy). In this model the coupling to the superconductor is given
by the matrix 
\begin{equation}
	\Gamma_{ij} =
	\left\{ \! \begin{array}{ll}
		\Gamma, & i=j, \\
		\gamma, & i\neq j. \\
	\end{array} \right.
	\label{G_ij_mod}
\end{equation}
The diagonal elements $\Gamma$ correspond to the regular Andreev reflection, and
off-diagonal elements $\gamma$ describe crossed Andreev reflection (CAR).
The situation $\gamma < \Gamma$ is possible, e.g., in the fork geometry
shown in Fig.~\ref{fig:fork}.

The results of the numerical calculation for $\varepsilon_1 = \varepsilon_2 =
\ldots = \varepsilon_{\scriptscriptstyle N}$ and $U = \Gamma$ are presented in
Figs.~\hyperref[fig:degeneracies_2ch]{\ref{fig:degeneracies_2ch}(a)}-\hyperref[fig:degeneracies_2ch]{\ref{fig:degeneracies_2ch}(c)} and
\hyperref[fig:degeneracies_2ch]{\ref{fig:degeneracies_2ch}(g)}-\hyperref[fig:degeneracies_2ch]{\ref{fig:degeneracies_2ch}(l)} [$N=2$] as well as
Figs.~\ref{fig:degeneracies_3ch} [$N=3$] and
Figs.~\hyperref[fig:degeneracies_4ch]{\ref{fig:degeneracies_4ch}(a)}-\hyperref[fig:degeneracies_4ch]{\ref{fig:degeneracies_4ch}(i)}. Here
${\tilde\varepsilon} = \varepsilon_1 + 2U + 2(N-1)u$.

In the absence of Coulomb interaction between the channels, $u = 0$, and zero
CAR $\gamma=0$, the channel contributions are additive and the system is
either in the nondegenerate state $\mathcal{P}_{\scriptscriptstyle N}=0$ or in
the state with $\mathcal{P}_{\scriptscriptstyle N}=N$ and degeneracy
$2^{\mathcal{P}_{\scriptscriptstyle N}}$, see
Figs.~\hyperref[fig:degeneracies_2ch]{\ref{fig:degeneracies_2ch}(a)}, \hyperref[fig:degeneracies_3ch]{\ref{fig:degeneracies_3ch}(a)}, and
\hyperref[fig:degeneracies_4ch]{\ref{fig:degeneracies_4ch}(a)}. A nonzero $\gamma$ leads to smearing of the
borders between $\mathcal{P}_{\scriptscriptstyle N}=0$ and
$\mathcal{P}_{\scriptscriptstyle N}=N$ regions, see
Figs.~\hyperref[fig:degeneracies_2ch]{\ref{fig:degeneracies_2ch}(b)}, \hyperref[fig:degeneracies_3ch]{\ref{fig:degeneracies_3ch}(b)}, and
\hyperref[fig:degeneracies_4ch]{\ref{fig:degeneracies_4ch}(b)} for $\gamma=0.1\Gamma$ and
Figs.~\hyperref[fig:degeneracies_2ch]{\ref{fig:degeneracies_2ch}(c)}, \hyperref[fig:degeneracies_3ch]{\ref{fig:degeneracies_3ch}(c)}, and
\hyperref[fig:degeneracies_4ch]{\ref{fig:degeneracies_4ch}(c)} for $\gamma=\Gamma$.

For $0 < u < U$ and $\gamma=0$, the areas with integer
$\mathcal{P}_{\scriptscriptstyle N}=0$, $1$, $\ldots$, $N$ are shown in
Figs.~\hyperref[fig:degeneracies_2ch]{\ref{fig:degeneracies_2ch}(g)}, \hyperref[fig:degeneracies_3ch]{\ref{fig:degeneracies_3ch}(d)}, and
\hyperref[fig:degeneracies_4ch]{\ref{fig:degeneracies_4ch}(d)}. The ``centers'' of the regions with
$\mathcal{P}_{\scriptscriptstyle N}>0$ are separated by distances defined by
the off-diagonal elements of $U_{ij}$, e.g., for Eq.~(\ref{U_ij_mod}) these
distances are $4u$. Nonzero values of $\gamma$ result in continuous 
$\mathcal{P}_{\scriptscriptstyle N}$ in $[0 \ldots N]$, see
Figs.~\hyperref[fig:degeneracies_2ch]{\ref{fig:degeneracies_2ch}(h)}-\hyperref[fig:degeneracies_2ch]{\ref{fig:degeneracies_2ch}(i)},
\hyperref[fig:degeneracies_3ch]{\ref{fig:degeneracies_3ch}(e)}-\hyperref[fig:degeneracies_3ch]{\ref{fig:degeneracies_3ch}(f)}, and
\hyperref[fig:degeneracies_4ch]{\ref{fig:degeneracies_4ch}(e)}-\hyperref[fig:degeneracies_4ch]{\ref{fig:degeneracies_4ch}(f)}.

For $u = U$ and $\gamma=0$ there are $N$ regions separated by $4u$ with
$\mathcal{P}_{\scriptscriptstyle N}=1$ [Figs.~\hyperref[fig:degeneracies_2ch]{\ref{fig:degeneracies_2ch}(j)},
\hyperref[fig:degeneracies_3ch]{\ref{fig:degeneracies_3ch}(g)}, and \hyperref[fig:degeneracies_4ch]{\ref{fig:degeneracies_4ch}(g)}]. CAR
$\gamma>0$ increases $\mathcal{P}_{\scriptscriptstyle N}$ in regions which
were zero initially at $\gamma=0$; $\mathcal{P}_{\scriptscriptstyle
N}$ can be increased as well as decreased from 1
[Figs.~\hyperref[fig:degeneracies_2ch]{\ref{fig:degeneracies_2ch}(k)}-\hyperref[fig:degeneracies_2ch]{\ref{fig:degeneracies_2ch}(l)},
\hyperref[fig:degeneracies_3ch]{\ref{fig:degeneracies_3ch}(h)}-\hyperref[fig:degeneracies_3ch]{\ref{fig:degeneracies_3ch}(i)}, and
\hyperref[fig:degeneracies_4ch]{\ref{fig:degeneracies_4ch}(h)}-\hyperref[fig:degeneracies_4ch]{\ref{fig:degeneracies_4ch}(i)}].

\subsection{Nanotubes}

In this section, we describe Andreev quantum dots based on a single-wall
nanotube, two parallel single-wall nanotubes [Fig.~\hyperref[fig:andreev_dot]{\ref{fig:andreev_dot}(b)}],
or a multiwall nanotube/molecule [Fig.~\hyperref[fig:andreev_dot]{\ref{fig:andreev_dot}(c)}].

The single-wall nanotube specifies a two-channel Coulomb matrix and an Andreev
reflection matrix
\begin{equation}
	U_{ij} = 
	\left[ \! \begin{array}{cc}
		U & U \\
		U & U \\
	\end{array} \! \right] \!, \quad
	\Gamma_{ij} = 
	\left[ \! \begin{array}{cc}
		\Gamma & \Gamma \\
		\Gamma & \Gamma \\
	\end{array} \! \right] \!.
	\label{UG_ij_SWNT}
\end{equation}
The existence of two orbital states leads to the appearance of two doubly
degenerate regions, see Fig.~\hyperref[fig:degeneracies_2ch]{\ref{fig:degeneracies_2ch}(l)}.

Two parallel single-wall nanotubes can be described by the Coulomb matrix
\begin{equation}
	U_{ij} =
	\left[ \! \begin{array}{cccc}
		U & U & u & u \\
		U & U & u & u \\
		u & u & U & U \\
		u & u & U & U
	\end{array} \! \right] \!
	\label{U_ij_2SWNT}
\end{equation}
and an AR matrix 
\begin{equation}
	\Gamma_{ij} = 
	\left[ \! \begin{array}{cccc}
		\Gamma & \Gamma & \gamma & \gamma \\
		\Gamma & \Gamma & \gamma & \gamma \\
		\gamma & \gamma & \Gamma & \Gamma \\
		\gamma & \gamma & \Gamma & \Gamma
	\end{array} \! \right] \!.
	\label{G_ij_2SWNT}
\end{equation}
The $2 \times 2$ block structure appears due to the twofold orbital degeneracy
in the single-wall nanotube, where $u$ and $\gamma$ describe the interaction
between nanotubes. The structure of the ground state for $\gamma=0$ is
presented in Fig.~\hyperref[fig:degeneracies_4ch]{\ref{fig:degeneracies_4ch}(j)}: both regions with
$\mathcal{P}_{\scriptscriptstyle N}=1$ (two-fold degeneracy) and
$\mathcal{P}_{\scriptscriptstyle N}=2$ (four-fold degeneracy) are present.
Two ``copies'' of 1-2-1 regions appear due to orbital degeneracy in each
nanotube, which can be compared to Fig.~\hyperref[fig:degeneracies_2ch]{\ref{fig:degeneracies_2ch}(g)} for the
case with no orbital degeneracy. In Figs.~\hyperref[fig:degeneracies_4ch]{\ref{fig:degeneracies_4ch}(k)} and
\hyperref[fig:degeneracies_4ch]{\ref{fig:degeneracies_4ch}(l)} the $\mathcal{P}_{\scriptscriptstyle N}$
behavior at $\gamma=0.1\Gamma$ and $\gamma=\Gamma$ is shown.

The multiwall nanotube has four or more orbital states. The most simple case,
$N=4$, can be described by the interaction matrix Eq.~(\ref{U_ij_2SWNT}) with
$u = U$ [or the same Eq.~(\ref{U_ij_mod})]. This situation is illustrated in
Figs.~\hyperref[fig:degeneracies_4ch]{\ref{fig:degeneracies_4ch}(g)}-\hyperref[fig:degeneracies_4ch]{\ref{fig:degeneracies_4ch}(i)}.

In the case of a diagonal $\Gamma_{ij}$ ($\Gamma_{ij}=0$, $i\neq j$), the
ground state is created from the singlet $|0\rangle$, $|2\rangle$ or doublet
$|\!\!\uparrow\rangle$, $|\!\!\downarrow\rangle$ states (but not a mixture of
singlet and doublet states of the same channel) and
$\mathcal{P}_{\scriptscriptstyle N}$ takes on integer values only. In the case
of arbitrary off-diagonal elements $\Gamma_{ij}$ ($\Gamma_{ij}\geqslant 0$,
$i\neq j$), the ground state can be a mixture of all possible states including
simultaneously $|0\rangle$, $|\!\!\uparrow\rangle$, $|\!\!\downarrow\rangle$,
and $|2\rangle$ states from the same channel.

The value of $\mathcal{P}_{\scriptscriptstyle N}$ has a large impact on the
physical properties of the dot, such as magnetic properties, charge on the
dot, and the Josephson current.

\section{Magnetic properties 
\label{sec:magnetic}}

In this section, we describe the magnetic properties of the ground state in the
presence of a weak external magnetic field. We take into account only the
Zeeman splitting and neglect all other lower-order effects such as Rashba
spin-orbit coupling (see Ref.~\onlinecite{Sun:2001} for more details). This
situation can be described by an additional term to the
Hamiltonian~(\ref{H_eff}) with a magnetic field $B$ in the transverse
direction
\begin{equation}
	\mathcal{\hat H}_{\rm\scriptscriptstyle M} = 
	\frac{g \mu_{\rm\scriptscriptstyle B} B}{\hbar}
	\sum\limits_{i=1}^N 
	\sum\limits_\sigma 
	s_\sigma {\hat n}_{i\sigma},
	\label{H_magnetic}
\end{equation}
where $g \approx 2$ is the Land\'e factor and $\mu_{\rm\scriptscriptstyle B} =
|e|\hbar / 2m$ the Bohr magneton. The spin quantum number is
$s_{\downarrow/\uparrow} = \mp \hbar/2$ in the corresponding $i^\text{th}$
channel. In the presence of a magnetic field, the doublet state of each
channel splits into two levels with energy difference $g
\mu_{\rm\scriptscriptstyle B} B$.

In the absence of CAR, $\mathcal{P}_{\scriptscriptstyle N}$ takes on only the
integer values from $0$ to $N$. Then, the lowest level of the system with spin
$-\mathcal{P}_{\scriptscriptstyle N}/2$ (where
$\mathcal{P}_{\scriptscriptstyle N}$ is the number of channels in this doublet
state) is nondegenerate, the first excited with spin
$-(\mathcal{P}_{\scriptscriptstyle N}-2)/2$ is
$\mathcal{P}_{\scriptscriptstyle N}$-degenerate, etc. The whole picture is
presented in Table~\ref{tab:Zeeman}. In this table
$C_{\mathcal{P}_{\scriptscriptstyle N}}^i \equiv
\mathcal{P}_{\scriptscriptstyle N}! / (\mathcal{P}_{\scriptscriptstyle
N}-i)!\,i!$ and naturally $\sum_{i=0}^{\mathcal{P}_{\scriptscriptstyle N}}
C_{\mathcal{P}_{\scriptscriptstyle N}}^i = 2^{\mathcal{P}_{\scriptscriptstyle
N}}$.
\begin{table}[h!] 
	\caption{
		The spins and degeneracies of the states obtained from a
		$2^{\mathcal{P}_{\scriptscriptstyle N}}$-degenerate ground state 
		with Zeeman splitting.
	}
	\begin{center}
	\begin{tabular}{lll}
		\hline \hline
		\; State \; & \; Spin \; & \; Degeneracy \; \\
		\hline
		\; $\uparrow \uparrow \ldots \uparrow \uparrow$ \; & \; ${\phantom -}\mathcal{P}_{\scriptscriptstyle N}/2$ \; & \; $C_{\mathcal{P}_{\scriptscriptstyle N}}^{\mathcal{P}_{\scriptscriptstyle N}} = 1$ \; \\
		\; $\uparrow \uparrow \ldots \uparrow \downarrow$ \; & \; ${\phantom -}\mathcal{P}_{\scriptscriptstyle N}/2-1$ \; & \; $C_{\mathcal{P}_{\scriptscriptstyle N}}^{\mathcal{P}-1} = \mathcal{P}_{\scriptscriptstyle N}$ \; \\
		\; $\cdots$ \; & \; $\cdots$ \; & \; $\cdots$ \; \\ 
		\; $\uparrow \uparrow \ldots \downarrow \downarrow$ \; & \; $-\mathcal{P}_{\scriptscriptstyle N}/2+i$ \; & \; $C_{\mathcal{P}_{\scriptscriptstyle N}}^i$ \; \\
		\; $\cdots$ \; & \; $\cdots$ \; & \; $\cdots$ \; \\ 
		\; $\uparrow \downarrow \ldots \downarrow \downarrow$ \; & \; $-\mathcal{P}_{\scriptscriptstyle N}/2+1$ \; & \; $C_{\mathcal{P}_{\scriptscriptstyle N}}^1 = \mathcal{P}_{\scriptscriptstyle N}$ \; \\
		\; $\downarrow \downarrow \ldots \downarrow \downarrow$ \; & \; $-\mathcal{P}_{\scriptscriptstyle N}/2$ \; & \; $C_{\mathcal{P}_{\scriptscriptstyle N}}^0 = 1$ \; \vspace{0.5mm} \\
		 \hline \hline
	\end{tabular}
	\end{center}
	\label{tab:Zeeman}
\end{table}
The magnetic properties of the system correspond to a system of
$\mathcal{P}_{\scriptscriptstyle N}$ independent spins $1/2$. The influence of
the Hamiltonian~(\ref{H_magnetic}) leads to a change in size of the degenerate
regions, which is equivalent to changing the Coulomb interaction $U$. The
dependence of $\mathcal{P}_{\scriptscriptstyle N}$ on~$\varphi$
and~$\tilde\varepsilon$ has been discussed in Sec.~\ref{sec:degeneracies}.

\section{Charge and Josephson current
\label{sec:charge_current}}

\begin{figure*}[tb]
	\includegraphics[width=12.5cm]{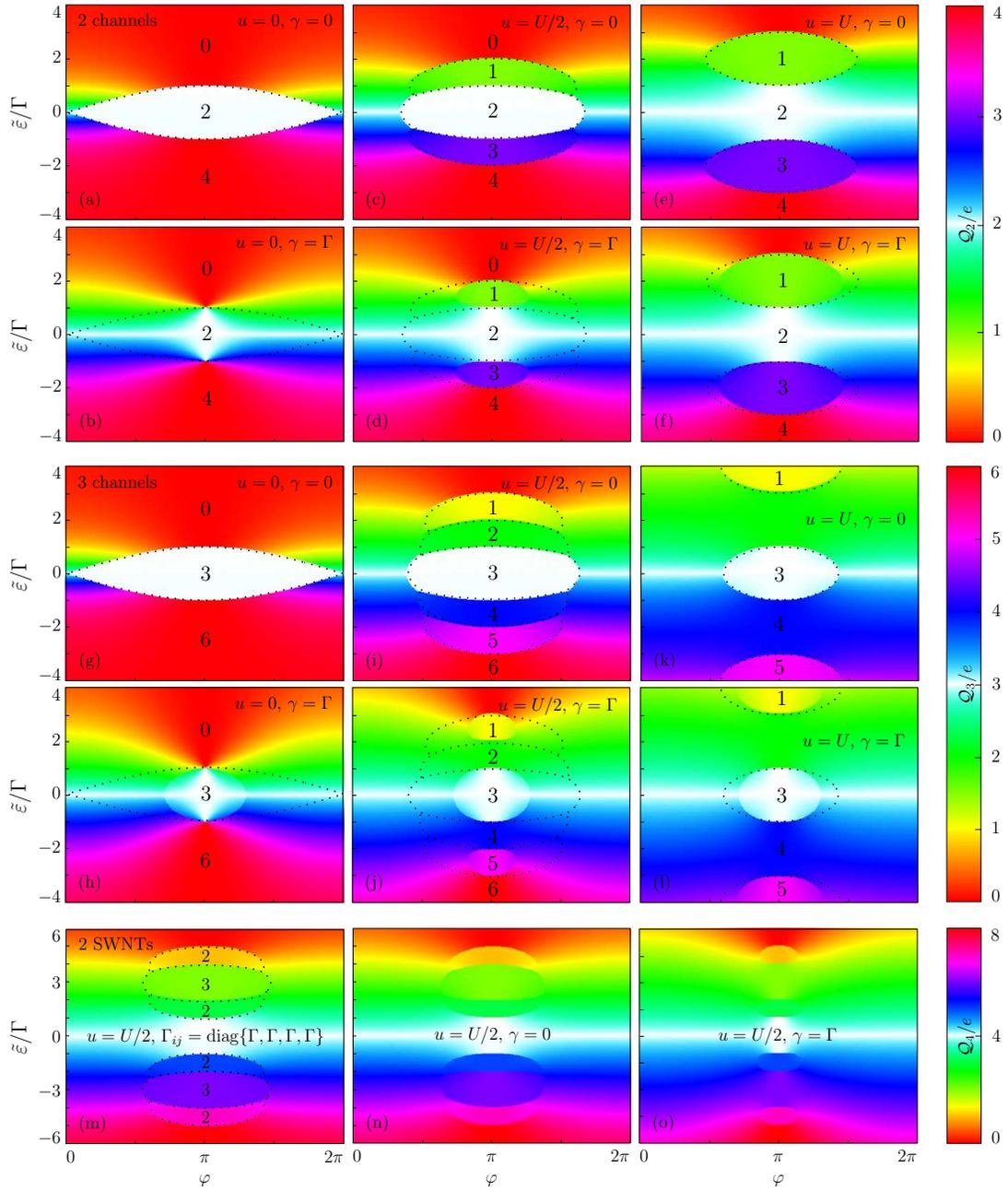}
	\caption{
(Color online) Charge of the ground state $\mathcal{Q}_{2,3,4}$ in $(\varphi,
{\tilde\varepsilon})$ space. Plots (a)-(f) [$N=2$], (g)-(l) [$N=3$] are for
$U_{ij}$ and $\Gamma_{ij}$ as defined by Eqs.~(\ref{U_ij_mod}) and
~(\ref{G_ij_mod}). Plot (m) corresponds to the two nanotubes [$U_{ij}$ defined
by Eq.~(\ref{U_ij_2SWNT})] and zero CAR between all channels
[Eq.~(\ref{G_ij_mod})]. Plots (n) and (o) describe two nanotubes [$U_{ij}$ and
$\Gamma_{ij}$ are given by Eqs.~(\ref{U_ij_2SWNT}) and (\ref{G_ij_2SWNT})]
with suppressed and maximal CAR between nanotubes. In all plots the charges
goes from 0 (cold red) to $2Ne$ (warm red); the ``centralized'' charge $Ne$ is
shown in white color. Note that for all different $N$ there are different
colorbars.
	}
	\label{fig:charges}
\end{figure*}

\begin{figure*}[tb]
	\includegraphics[width=12.85cm]{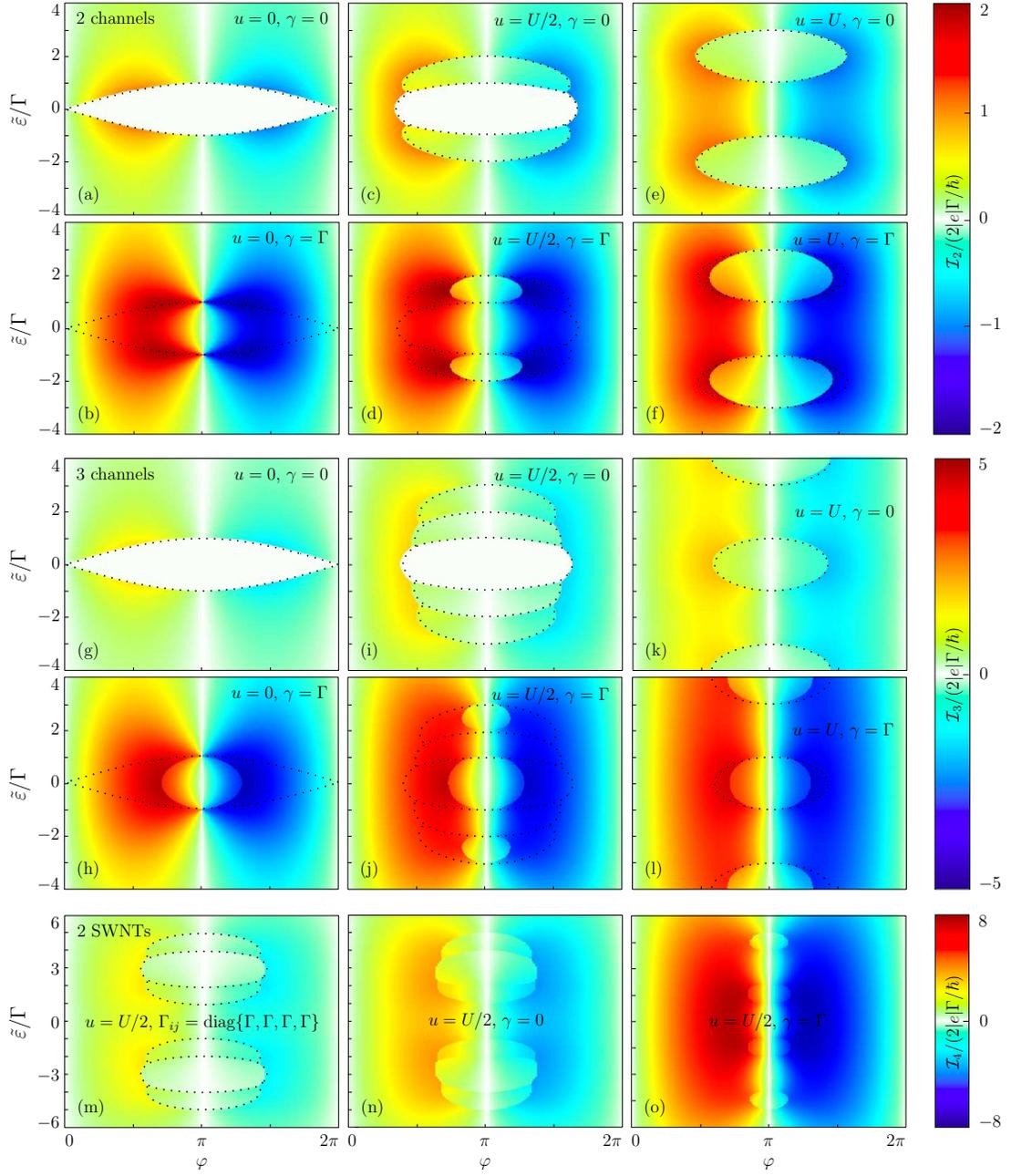}
	\caption{
(Color online) The Josephson current of the ground state $\mathcal{I}_{2,3,4}$ in $(\varphi, {\tilde\varepsilon})$ space. All parameters are the same as in Fig.~\ref{fig:charges}. A zero current is depicted by white color, negative by blue, and positive by red. For all $N=2,3,4$ there are different colorbars.
	}
	\label{fig:currents}
\end{figure*}

The operator for the total dot charge is given by
\begin{equation}
	\mathcal{\hat Q}_{\scriptscriptstyle N} =
	\sum\limits_{i=1}^N 
	\mathbbm{1}_1 \otimes \ldots \otimes {\hat Q}_i \otimes \ldots
	\otimes\mathbbm{1}_{\scriptscriptstyle N},
	\label{dot_Q}
\end{equation}
where ${\hat Q}_i = e {\hat n}_i = e \, ({\hat d}_{i\uparrow}^\dag {\hat
d}_{i\uparrow}^{\phantom\dag} + {\hat d}_{i \downarrow}^\dag {\hat
d}_{i\downarrow}^{\phantom\dag})$ is the $i^\text{th}$ channel charge
operator. The latter has matrix elements $Q_i = e \, {\rm diag}_i \{0, 1, 1,
2\}$. The dot charge~(\ref{dot_Q}) has matrix elements
\begin{align}
	\mathcal{Q}_{\scriptscriptstyle N} =
	e \sum\limits_{i=1}^N n_i,
	\label{dot_Q_me}
\end{align}
where the matrices $n_i$ are inserted using the same rules as in
formula~(\ref{H_eff_N_me}).

The charge of the ground state is presented in Fig.~\ref{fig:charges} as a
function of $\varphi$ and ${\tilde\varepsilon}$. In each channel, the charge
can vary from $2e$ to $0$ when increasing ${\tilde\varepsilon}$; the Coulomb
interaction $u$ and crossed Andreev reflections $\gamma$ change the total
charge from the straightforward sum of channel charges.

In the single channel case $N=1$, the charge $\mathcal{Q}_1$ can vary from $0$
to $2e$; for the diagonal $U$ and $\Gamma$ matrices [Eqs.~(\ref{U_ij_mod}) and
(\ref{G_ij_mod})], the total charge $\mathcal{Q}_{2,3}$ is reduced to the sum
of charges from each channel as shown for $N=2$ in Fig.~\hyperref[fig:charges]{\ref{fig:charges}(a)}
and for $N=3$ in Fig.~\hyperref[fig:charges]{\ref{fig:charges}(g)}. The white flat plateau in the
middle part of the plot corresponds to the region with
$\mathcal{P}_{\scriptscriptstyle N} = N$ and $\mathcal{Q}_{\scriptscriptstyle
N} = eN$; in this region, the charge does not depend on $\varphi$ or
${\tilde\varepsilon}$.\cite{Sadovskyy:2007b,Sadovskyy:2010} The nonzero $u$
``divides'' this region into pieces: partially [e.g.
Figs.~\hyperref[fig:charges]{\ref{fig:charges}(c)} and~\hyperref[fig:charges]{\ref{fig:charges}(i)}] or fully [e.g.
Figs.~\hyperref[fig:charges]{\ref{fig:charges}(e)} and~\hyperref[fig:charges]{\ref{fig:charges}(k)}]. In the regions with
$\mathcal{P}_{\scriptscriptstyle N} < N$, only charges from $N -
\mathcal{P}_{\scriptscriptstyle N}$ channels contribute to the phase-dependent
part of the total charge; the remaining channels
($\mathcal{P}_{\scriptscriptstyle N}$) give a constant contribution $e$. As
previously, CAR smears the borders of the plateau as shown in
Figs.~\hyperref[fig:charges]{\ref{fig:charges}(b)}-\hyperref[fig:charges]{\ref{fig:charges}(f)} and
\hyperref[fig:charges]{\ref{fig:charges}(h)}-\hyperref[fig:charges]{\ref{fig:charges}(l)}. In
Figs.~\hyperref[fig:charges]{\ref{fig:charges}(m)}-\hyperref[fig:charges]{\ref{fig:charges}(o)} the total charge
$\mathcal{Q}_4$ of the two SWNTs connected in parallel is presented; for
comparison see
Figs.~\hyperref[fig:degeneracies_4ch]{\ref{fig:degeneracies_4ch}(j)}-\hyperref[fig:degeneracies_4ch]{\ref{fig:degeneracies_4ch}(l)} with
$\mathcal{P}_4$. The charge in each channel is changed from $0$ to $2e$, so
the total charge goes from $0$ to $2eN$. The scaling of the phase-dependent
part of the charge will be analyzed in Sec.~\ref{sec:sensitivity}.

The total Josephson current operator is given by the expression
\begin{equation}
	\mathcal{\hat I}_{\scriptscriptstyle N} =
	\sum\limits_{i=1}^N 
	\mathbbm{1}_1 \otimes \ldots \otimes {\hat I}_i \otimes \ldots
	\otimes\mathbbm{1}_{\scriptscriptstyle N},
	\label{dot_I}
\end{equation}
where 
\begin{equation}
	{\hat I}_i
	= -\frac{2e}{\hbar} \, \Gamma_i \sin\frac{\varphi}{2} \, 
	(
	{\hat d}_{i\downarrow}^{\phantom\dag} {\hat d}_{i\uparrow}^{\phantom\dag} + 
	{\hat d}_{i\uparrow}^\dag {\hat d}_{i\downarrow}^\dag
	)
	\label{I_op}
\end{equation}
is the $i^\text{th}$ channel current operator. The latter has matrix elements 
\begin{equation}
	I_i = -\frac{2e}{\hbar} \, \Gamma_i \sin\frac{\varphi}{2} \,
	{\rm adiag}_i \{1, 0, 0, 1\}.
	\label{I_me}
\end{equation}
The total current operator can be generated in the same way as in
Eqs.~(\ref{H_eff_N_me}) and (\ref{dot_Q_me}). Note that the charge
$\mathcal{\hat I}_{\scriptscriptstyle N}$ and current $\mathcal{\hat
I}_{\scriptscriptstyle N}$ can be obtained as a derivate with respect to
$V_{\rm g}$ and $\varphi$, respectively. The Josephson current
$\mathcal{I}_{\scriptscriptstyle N}$ corresponding to the
operator~(\ref{dot_I}) is presented in Fig.~\ref{fig:currents} as a function
of the superconducting phase difference $\varphi$ and ${\tilde\varepsilon}$.

In the noninteracting case $u=0$ and $\gamma=0$, the current is defined by
energy the levels [Eqs.~(\ref{S_en}) and (\ref{D_en})] and scales proportional to
the number of channels $N$, see Fig.~\hyperref[fig:currents]{\ref{fig:currents}(a)} for $N=2$ and in
Fig.~\hyperref[fig:currents]{\ref{fig:currents}(g)} for $N=3$. The white region corresponds to the
zero-current region; along the horizontal lines the current demonstrates the
usual sine behavior (if the channel is in the singlet state) or zero current
behavior (if the channel is in the doublet
state).\cite{Rozhkov:2000,Sadovskyy:2010} In the maximally degenerate regions
$\mathcal{P}_{\scriptscriptstyle N} = N$, the current can flow in the opposite
direction ($\pi$-junction) if the continuous spectrum above $\Delta$ is taken
into account.\cite{Rozhkov:2000,Vecino:2003,Dam:2006,Zazunov:2009,Zazunov:2010} 
The channels can be in the doublet state simultaneously [Figs.~\hyperref[fig:currents]{\ref{fig:currents}(a)} 
and \hyperref[fig:currents]{\ref{fig:currents}(g)}] or asynchronously [Figs.~\hyperref[fig:currents]{\ref{fig:currents}(c)}, \hyperref[fig:currents]{\ref{fig:currents}(e)}, 
\hyperref[fig:currents]{\ref{fig:currents}(i)}, and \hyperref[fig:currents]{\ref{fig:currents}(k)}]. The small CAR $\gamma$ smears 
the borders of the degenerate regions [Figs.~\hyperref[fig:currents]{\ref{fig:currents}(b)}, \hyperref[fig:currents]{\ref{fig:currents}(d)}, 
and \hyperref[fig:currents]{\ref{fig:currents}(f)}]; large CAR decreases or even eliminates the degenerate
regions [Figs.~\hyperref[fig:currents]{\ref{fig:currents}(h)}, \hyperref[fig:currents]{\ref{fig:currents}(j)}, and \hyperref[fig:currents]{\ref{fig:currents}(l)}].

The total current through two parallel SWNTs is shown in
Figs.~\hyperref[fig:currents]{\ref{fig:currents}(m)}-\hyperref[fig:currents]{\ref{fig:currents}(o)}; for comparison see
Figs.~\hyperref[fig:degeneracies_4ch]{\ref{fig:degeneracies_4ch}(j)}-\hyperref[fig:degeneracies_4ch]{\ref{fig:degeneracies_4ch}(l)}.

The Coulomb interaction as usually suppresses the current; CAR adds additional
$N(N-1)$ channels for the current and therefore increases the current. Note
that in the limits considered here, the maximal $\mathcal{P}_{\scriptscriptstyle N}
= N$ leads to zero charge and current response with respect to $\varphi$; with
decreasing $\mathcal{P}_{\scriptscriptstyle N}$ the response increases.

\section{Sensitivity
\label{sec:sensitivity}}

\begin{figure*}[tb]
	\includegraphics[width=12.99cm]{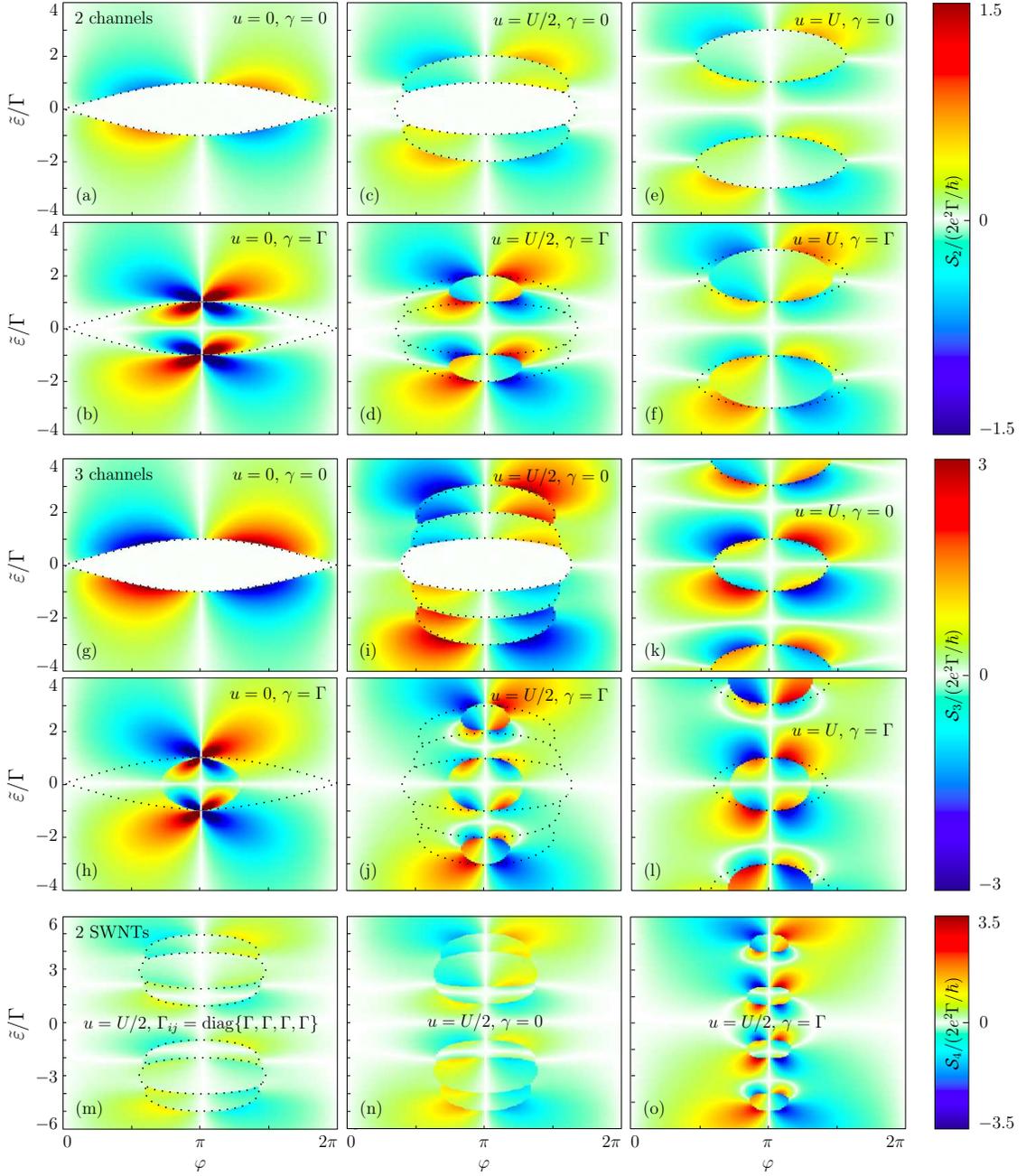}
	\caption{
(Color online) Charge-to-phase sensitivity in the ground state $\mathcal{I}_{2,3,4}$ in $(\varphi, {\tilde\varepsilon})$ space. All parameters coincide with Figs.~\ref{fig:charges} and \ref{fig:currents}. In the plots (b) and (h) the sensitivity diverges in $(\varphi, {\tilde\varepsilon}) = (\pi, \pm\Gamma)$ and sensitivity maximum (minimum) was cut by hands. For all $N$ there are a different colorbars.
	}
	\label{fig:sensitivities}
\end{figure*}

Let us now analyze the sensitivity of the charge $\mathcal{Q}$ with the
superconducting phase difference $\varphi$. The most relevant physical
quantity is the differential sensitivity $\mathcal{S} = (4e/\hbar)
\partial_\varphi \mathcal{Q}$. It characterizes the charge response to
infinitely small deviations of the superconducting phase and describes the
operating component in new types of magnetometers based on the charge 
of the Andreev quantum dot proposed in Refs.~\onlinecite{Sadovskyy:2007b}
and \onlinecite{Sadovskyy:2010}. The above defined sensitivity coincides with 
the current-to-gate voltage sensitivity $\mathcal{\tilde S} = e \, \partial \mathcal{I} 
/ \partial {\tilde\varepsilon}$, which characterizes the Josephson
transistor.\cite{Kuhn:2001} This correspondence can easily be proven if we
remember that the charge is the energy derivative with respect to the gate
voltage, and the current is the energy derivative with respect to the phase.
This implies that the results we find for the differential sensitivity of the
magnetometer are also true for the sensitivity of the Josephson transistor. We
will denote $\mathcal{S}_{\scriptscriptstyle N}$ as the sensitivity of the
system with $N$ channels.

The sensitivity as a function of the phase difference $\varphi$ and the gate
voltage ${\tilde\varepsilon}$ is displayed in Fig.~\ref{fig:sensitivities}.
Given the case of multiple channels and the absence of Coulomb interaction
between them, it is natural to expect the sensitivity to be composed from the
sensitivity of each channel separately as shown in
Fig.~\hyperref[fig:sensitivities]{\ref{fig:sensitivities}(a)} and in Fig.~\hyperref[fig:sensitivities]{\ref{fig:sensitivities}(g)}. In
this section we examine the influence of the Coulomb interaction between
channels and CAR on the sensitivity.

Due to the slow singlet-doublet transitions, we analyze only the sensitivity
which comes from the channels in the singlet regime and omit the sensitivity
given by the step in the charge during the singlet-doublet transition: For 
low-frequency measurements this kind of sensitivity may be quite important.

In the absence of the Coulomb interaction $u=U=0$ and the same normal dot
levels $\varepsilon_1 = \varepsilon_2 = \ldots =
\varepsilon_{\scriptscriptstyle N}$, the differential sensitivity has a maximum
near point $\varphi=\pi$ and ${\tilde\varepsilon} = \varepsilon_1 =
0$.\cite{Sadovskyy:2007b}

In case of a single channel $N=1$ with Coulomb interaction $U = \Gamma$,
the charge at $\varphi = 0$ and $\varphi = \pi$ is insensitive to the phase.
The maximal sensitivity appears at the border of the doublet region. See the
shape Fig.~\hyperref[fig:sensitivities]{\ref{fig:sensitivities}(a)} for $N=2$ and
Fig.~\hyperref[fig:sensitivities]{\ref{fig:sensitivities}(g)} for $N=3$.

If we consider the two-channel case $N=2$ with different dot levels
$\varepsilon_1 \neq \varepsilon_2$ without the Coulomb interaction $u=U=0$,
then the sensitivity is defined by the distance $\delta\varepsilon_{12} =
\varepsilon_2 - \varepsilon_1$. If this distance equals to zero, then the
sensitivity $\mathcal{S}_2 = 2\mathcal{S}_1$. With increasing
$|\delta\varepsilon_{12}|$ the sensitivities associated with different
channels are canceled inside the region $[\varepsilon_1 \ldots \varepsilon_2]$
due to the different signs in the charge of the first and the second level and
partially added outside this region. Given the large level separation
$|\delta\varepsilon_{12}| \gg \Gamma$, the maximum sensitivity drops to the
sensitivity of the single channel case $\mathcal{S}_2 = \mathcal{S}_1$.

The effects of the Coulomb interaction in the multichannel case $N>1$ may
be separated into the effect of degenerate regions ($0 <
\mathcal{P}_{\scriptscriptstyle N} < N$ or $\mathcal{P}_{\scriptscriptstyle N}
= N$), which are either partially or totally insensitive to the phase, and the
effect of the repulsion between normal dot levels. Note that the size of the
regions with $\mathcal{P}_{\scriptscriptstyle N} > 0$ decreases with
increasing Coulomb interactions between levels, which can be seen by
comparing Figs.~\hyperref[fig:sensitivities]{\ref{fig:sensitivities}(a)}, \hyperref[fig:sensitivities]{\ref{fig:sensitivities}(c)},
and~\hyperref[fig:sensitivities]{\ref{fig:sensitivities}(e)} or Figs.~\hyperref[fig:sensitivities]{\ref{fig:sensitivities}(g)},
\hyperref[fig:sensitivities]{\ref{fig:sensitivities}(i)}, and~\hyperref[fig:sensitivities]{\ref{fig:sensitivities}(k)}. The combination of
the normal level repulsion and the emergence of the insensitive regions leads
to ``oscillations'' in the sensitivity as a function of $\varphi$ and
$\tilde\varepsilon$ as presented by the blue and red color in
Fig.~\ref{fig:sensitivities}.

The sensitivity as shown in Figs.~\hyperref[fig:sensitivities]{\ref{fig:sensitivities}(b)}, \hyperref[fig:sensitivities]{\ref{fig:sensitivities}(d)}, 
\hyperref[fig:sensitivities]{\ref{fig:sensitivities}(f)}, \hyperref[fig:sensitivities]{\ref{fig:sensitivities}(h)}, \hyperref[fig:sensitivities]{\ref{fig:sensitivities}(j)}, 
and \hyperref[fig:sensitivities]{\ref{fig:sensitivities}(l)} can either increase or decrease due to the
CAR. Partially, the CAR leads to a divergence in sensitivity at the points $(\varphi,
{\tilde\varepsilon}) = (\pi, \pm\Gamma)$ in Figs.~\hyperref[fig:sensitivities]{\ref{fig:sensitivities}(b)}
and \hyperref[fig:sensitivities]{\ref{fig:sensitivities}(h)}.

The sensitivity of the two parallel SWNTs is shown in
Figs.~\hyperref[fig:sensitivities]{\ref{fig:sensitivities}(m)}-\hyperref[fig:sensitivities]{\ref{fig:sensitivities}(o)}.

\section{Conclusion
\label{sec:conclusion}}

In this article we have described an Andreev quantum dot with several normal
levels/conducting channels in the infinite superconducting-gap limit. The
scaling of the charge and the current with the number of channels is the
central question of the work. We have introduced a recursive scheme for an
$N$ channel Hamiltonian and have analyzed it numerically/analytically.

This approach has allowed us to specify the degeneracy of the ground state,
depending on the superconducting phase difference and the position of the back
gate. Due to the Coulomb interaction inside each channel, doubly degenerate
ground states appear. For the case of the intermediate Coulomb interaction
between the channels, regions with a higher degeneracy are generated.
Nevertheless, when increasing the Coulomb interaction between the channels,
the doubly degenerate regions ``repel'' each other, and regions with higher
degeneracy disappear. The size of the doublet regions decrease due to the
Coulomb interaction between channels. The crossed Andreev reflection smears
the borders between regions with different degeneracy and can completely 
destroy the regions with higher degeneracy. Also, we have studied the magnetic
properties of the degenerate ground states in the presence of a Zeeman
splitting.

Finally, the charge of the intermediate region and the current of the
multichannel Andreev dot have been computed. The interplay between the
scaling and interaction effects has been discussed for some realistic
situations. The charge-to-phase and current-to-gate voltage sensitivities
increase with the number of channels but do not scale linearly as in the case
of independent channels. While the sensitivity always increases with the 
number of channels, the Coulomb interaction between channels leads 
to a sensitivity reduction. A similar behavior has been detected for the critical
current as a function of the number of channels. The multichannel device
could therefore be used as the sensitive magnetic flux detector or
alternatively as the Josephson transistor. Summing up, the charge, the
current, and the sensitivity scales linearly in the absence of crossed Andreev
reflections and Coulomb interaction between channels. The Coulomb interaction
suppresses the phase-dependent part, and CAR smears the jumps as a function of
phase difference and gate voltage.

\section*{Acknowledgments}

We acknowledge financial support by the CNRS LIA agreements with Landau
Institute, the RFBR Grant No.~11-02-00744-a (G.B.L.) and Grants Nos.~NSF
ECS-0608842, ARO W911NF-09-1-0395, and DARPA HR0011-09-1-0009
(I.A.S.).


\begin{thebibliography}{99}

\bibitem{Josephson:1962}
	B.D.~Josephson, 
	Phys.~Lett. {\bf 1}, 251 (1962).

\bibitem{Gennes:1964}
	P.G.~de~Gennes, 
	Rev.~Mod.~Phys. {\bf 36}, 225 (1964).

\bibitem{Anderson:1963}
	P.W.~Anderson and J.M.~Rowell,
	Phys.~Rev.~Lett. {\bf 10}, 230 (1963).

\bibitem{JarilloHerrero:2006}
	P.~Jarillo-Herrero, J.A.~van Dam, and L.P.~Kouwenhoven,
	Nature (London) {\bf 439}, 953 (2006).

\bibitem{Cleuziou:2006}
	J.-P.~Cleuziou, W.~Wernsdorfer, V.~Bouchiat, 
	T.~ Ondar\c{o}uhu, and M.~Monthioux,
	Nat. Nanotech. {\bf 1}, 53 (2006).

\bibitem{Chtchelkatchev:2003}
	N.M.~Chtchelkatchev and Yu.V.~Nazarov,
	Phys.~Rev.~Lett. {\bf 90}, 226806 (2003).

\bibitem{Sadovskyy:2007}
	I.A.~Sadovskyy, G.B.~Lesovik, and G.~Blatter, 
	Phys.~Rev.~B {\bf 75}, 195334 (2007).

\bibitem{Engstrom:2004}
	K.~Engstr\"om and J.~Kinaret, 
	Phys. Scr. {\bf 70}, 326 (2004). 

\bibitem{Rozhkov:2000}
	A.V.~Rozhkov and D.P.~Arovas,
	Phys.~Rev.~B {\bf 62}, 6687 (2000).

\bibitem{Sadovskyy:2007b}
	I.A.~Sadovskyy, G.B.~Lesovik, and G.~Blatter, 
	Pis'ma v ZhETF {\bf 86}, 239 (2007)
	[JETP Lett. {\bf 86}, 210 (2007)].

\bibitem{Saito:1998}
	R.~Saito, G.~Dresselhaus, and M.S.~Dresselhaus,
	{\it Physical Properties of Carbon Nanotubes}
	(Imperial College Press, London, 1998).

\bibitem{Falko:1992}
	V.I.~Falko and G.B.~Lesovik, 
	Solid~State~Comm. {\bf 84}, 835 (1992).

\bibitem{Andreev:1964a}
	A.F.~Andreev, 
	Zh.~Eksp.~Teor.~Fiz. {\bf 46}, 1823 (1964)
	[Sov.~Phys.~JETP {\bf 19}, 1228 (1964)].

\bibitem{Deutscher:2000}
	G.~Deutscher and D.~Feinberg,
	Appl.~Phys.~Lett. {\bf 76}, 81 (2000).

\bibitem{Beckmann:2004}
	D.~Beckmann, H.B.~Weber, and H.~v.~L\"ohneysen,
	Phys.~Rev.~Lett. {\bf 93}, 197003 (2004).

\bibitem{Beckmann:2006}
	D.~Beckmann and H.~v.~L\"ohneysen,
	AIP~Conf.~Proc. {\bf 850}, 875 (2006).

\bibitem{Lesovik:2001}
	G.B.~Lesovik, T.~Martin, and G.~Blatter, 
	Eur. Phys. J. B {\bf 24}, 287 (2001).

\bibitem{Kuhn:2001}
	D.D.~Kuhn, N.M.~Chtchelkatchev, G.B.~Lesovik, and G.~Blatter, 
	Phys.~Rev.~B {\bf 63}, 054520 (2001).

\bibitem{Chtchelkatchev:2000}
	N.M.~Chtchelkatchev, G.B.~Lesovik, and G.~Blatter,
	Phys.~Rev.~B {\bf 62}, 3559 (2000).

\bibitem{Landau:1984}
	L.D.~Landau, E.M.~Lifshitz, and L.P.~Pitaevskii, 
	{\it Electrodynamics of Continuous Media, Volume 8} (Pergamon Press, Oxford 1984).

\bibitem{Zazunov:2006}
	A.~Zazunov, D.~Feinberg, and T.~Martin, 
	Phys.~Rev.~Lett. {\bf 97}, 196801 (2006).

\bibitem{Sadovskyy:2010}
	I.A.~Sadovskyy, G.B.~Lesovik, T.~Jonckheere, and T.~Martin,
	Phys.~Rev.~B {\bf 82}, 235310 (2010).

\bibitem{MartinRodero:2011}
	A.~Mart\'in-Rodero and A.~Levy~Yeyati, 
	Adv.~Phys. {\bf 60}, 899 (2011).

\bibitem{Kubala:2003}
	B.~Kubala and J.~K\"onig, 
	Phys.~Rev.~B {\bf 67}, 205303 (2003).

\bibitem{Sun:2001}
	Q.F.~Sun, J.~Wang, and H.~Guo,
	Phys.~Rev.~B {\bf 71}, 165310 (2005).
 
\bibitem{Vecino:2003}
	E.~Vecino, A.~Mart\'in-Rodero, and A.~Levy~Yeyati,
	Phys.~Rev.~B {\bf 68} 035105 (2003).

\bibitem{Dam:2006}
	J.A.~van~Dam, Yu.V.~Nazarov, E.P.A.M.~Bakkers, 
	S.~De~Franceschi, and L.P.~Kouwenhoven,
	Nature (London) {\bf 442}, 667 (2006).

\bibitem{Zazunov:2009}
	A.~Zazunov, A.~Schulz, and R.~Egger,
	Phys.~Rev.~Lett. {\bf 102}, 047002 (2009).

\bibitem{Zazunov:2010}
	A.~Zazunov, A.~Levy~Yeyati, and R.~Egger,
	Phys.~Rev.~B {\bf 81}, 012502 (2010).

\end{thebibliography}
\end{document}